\documentclass[11pt]{article}
\usepackage[a4paper, margin = 1in]{geometry}
\usepackage[sorting = anyt, maxbibnames=9,maxcitenames=2, backend = bibtex, style=authoryear, doi=false, url=false, dashed=false]{biblatex}
\renewbibmacro{in:}{}

\usepackage[titletoc,page]{appendix}
\renewcommand*\appendixpagename{Appendix}
\renewcommand*\appendixtocname{Appendix}
\usepackage{graphicx}
\usepackage{setspace}
\usepackage{gensymb}
\usepackage{enumerate}
\usepackage{abstract,lipsum}
\usepackage[colorlinks=true,linkcolor=blue,citecolor=blue,urlcolor=blue]{hyperref}

\usepackage{amsmath}
\usepackage{amssymb}
\usepackage{amsthm}
\usepackage{amsfonts}
\usepackage{upgreek}
\usepackage{mathtools}
\usepackage{mathrsfs}
\usepackage{bm}

\usepackage[symbol]{footmisc}

\usepackage{color}
\usepackage{multirow}
\usepackage{algorithm,algorithmic}
\usepackage{verbatim}

\usepackage{tcolorbox}
\usepackage[english]{babel}
\theoremstyle{remark}

\usepackage{chngcntr} 

\begin{document}


\begin{center}
       \fontsize{18pt}{18pt}\selectfont \textbf{Extreme resilience and dissipation in heterogeneous disordered materials}
       
       \vspace*{0.3in}
       \fontsize{10pt}{10pt}\selectfont Jehoon Moon$^{1}$, Gisoo Lee$^{1}$, Jaehee Lee$^{1}$, Hansohl Cho$^{1,\dagger}$,  \\
       \vspace*{0.3in}
       \fontsize{10pt}{10pt}\selectfont $^{1}$Department of Aerospace Engineering, Korea Advanced Institute of Science and Technology, Daejeon, 34141, Republic of Korea \\ 

\vspace*{0.2in}
\fontsize{10pt}{10pt}\selectfont E-mails: $^\dagger$hansohl@kaist.ac.kr

\end{center}

\renewenvironment{abstract}
{\small 
\noindent \rule{\linewidth}{.5pt}\par{\noindent \bfseries \abstractname.}}
{\medskip\noindent \rule{\linewidth}{.5pt}
}

\vspace*{0.3in}
\onehalfspacing
\begin{abstract}
\fontsize{10pt}{10pt}\selectfont
Long range order and symmetry in heterogeneous materials architected on crystal lattices lead to elastic and inelastic anisotropies and thus limit mechanical functionalities in particular crystallographic directions. Here, we present a facile approach for designing heterogeneous disordered materials that exhibit nearly isotropic mechanical resilience and energy dissipation capabilities. We demonstrate, through experiments and numerical simulations on 3D-printed prototypes, that near-complete isotropy can be attained in the proposed heterogeneous materials with a small, finite number of random spatial points. We also show that adding connectivity between random subdomains leads to much enhanced elastic stiffness, plastic strength, energy dissipation, shape recovery, structural stability and reusability in our new heterogeneous materials. Overall, our study opens avenues for the rational design of a new class of heterogeneous materials with isotropic mechanical functionalities for which the engineered disorder throughout the subdomains plays a crucial role.
\\
\noindent
\end{abstract}

\doublespacing
\section{Introduction}
Heterogeneous materials composed of multiple dissimilar constituents that exhibit distinct physical properties are ubiquitous. Examples include block copolymers, metallic alloys and synthetic and natural composites that have been exploited for a myriad of engineering and biological applications. Of recent interest are mechanically resilient yet energy dissipative, heterogeneous materials often inspired by block copolymers (\cite{bates1990block, bockstaller2005block, qi2005stress, rinaldi2011tunable, cho2013dissipation, cho2017deformation}) and natural composites (\cite{munch2008tough, meyers2008biological, wegst2015bioinspired}) comprising plastomeric “hard” and elastomeric “soft” constituents thermodynamically immiscible. These naturally formed or man-made resilient yet dissipative materials are finding new avenues for multi-functional architectures and devices capable of withstanding mechanical, thermal and chemical extremes. Furthermore, recent advances in the additive manufacturing or three-dimensional printing of multiple materials have enabled the development and demonstration of heterogeneous materials that possess complex prescribed geometries and topologies from the sub-micrometer (\cite{lee2012periodic, zhang20203d,bauer2022nanoarchitected}) to the meter scale (\cite{wang2016lightweight, jia2019biomimetic, mueller2022architected,kotikian2024liquid}). 

Although heterogeneous materials with embedded intricate substructures have received great attention given their promising roles in achieving unique combinations of mechanical resilience and energy dissipation, most of the materials proposed thus far have been designed and demonstrated on simple crystalline (\cite{schaedler2011ultralight, jang2013fabrication,bauer2016approaching,meza2017reexamining,bonatti2017large}) or quasi-crystalline lattices (\cite{wang2020quasiperiodic,imediegwu2023mechanical,rosa2024stiff}). Albeit simple, the long-range order and symmetry in these heterogeneous materials constructed on crystalline lattices lead to intrinsic anisotropies in both elasticity and inelasticity, undesirable in many application domains. Recently, it has been demonstrated that combining several lattice materials that exhibit inverse anisotropies enables the rational design of heterogeneous materials that exhibit isotropic elasticity and inelasticity (\cite{berger2017mechanical,tancogne20183d,tancogne2018elastically,lee2024extreme}). Furthermore, random arrangements of constituent hard particles or fillers within a surrounding soft matrix have led to isotropic mechanical responses, especially under small strain levels (\cite{hashin1963variational, gusev1997representative, moulinec1998numerical, sheng2004multiscale}). However, in existing frameworks for these heterogeneous materials with random microstructures, it has been found that a large number of particles would be needed to reach near-complete mechanical isotropy, also dependent strongly upon the target volume fractions of the constituent materials.

Here, we have designed, simulated, fabricated and demonstrated a new class of highly resilient yet dissipative heterogeneous disordered materials with plastomeric hard and elastomeric soft components. We start by constructing the two-phase materials with a random dispersion of hard particles. Importantly, we show that connectivity between the random microstructures leads to a significant reduction in elastic and inelastic anisotropies. We demonstrate that near-complete isotropy (with a universal anisotropy index of $<$ 0.012) emerges in the heterogeneous disordered materials with a small, finite number of random spatial points, herein N = 7, clearly underpinned by the microstructural analysis of local symmetries. The influence of connectivity throughout the random microstructures is explored further for 3D-printed prototypes subjected to multiple loading and unloading cycles; quantitative agreement is found between experimental and numerical results, demonstrating that extreme resilience and energy dissipation independent of the loading direction can be achieved in these heterogeneous disordered materials by harnessing the connectivity.

The paper is organized as follows. A facile design procedure to construct the heterogeneous disordered materials is presented in Section \ref{section:Construt}. We then analyze the elastic anisotropy in these materials with (1) dispersed-particle and (2) continuous morphologies in Section \ref{sec:elasticAnisotropy}, which is further supported by microstructural and geometric details in statistical realizations for each of the representative volume elements with varying number of random spatial points in Section \ref{sec:Microstructural}. Large strain elastic and inelastic features in both experiments and numerical simulations are presented in Section \ref{sec:LargeStrain}; then, energy dissipation and load transfer capabilities are analyzed upon multiple cyclic loading, unloading, recovery and reloading scenarios in Section \ref{sec:Cyclic}, by which we further discuss reusability in our heterogeneous disordered materials with and without connectivity throughout the random subdomains. We close with concluding remarks on our findings and the main contributions of this work in Section \ref{sec:Consclusion}. Additionally, in Appendix \ref{Appendix:Microstructural}, the microstructural and geometric details in these materials are further provided. Detailed experimental, computational, and micromechanical modeling procedures to characterize the elastic and inelastic features are presented in Appendix \ref{Appendix:Experimental}. We finally present further analysis results on large strain isotropy in our heterogeneous disordered materials with continuous morphology in Appendix \ref{sec:inelastic_isotropy}.

\section{Constructing heterogeneous disordered materials}
\label{section:Construt}
\begin{figure}[h!]
    \centering
    \includegraphics[width=1.0\textwidth]{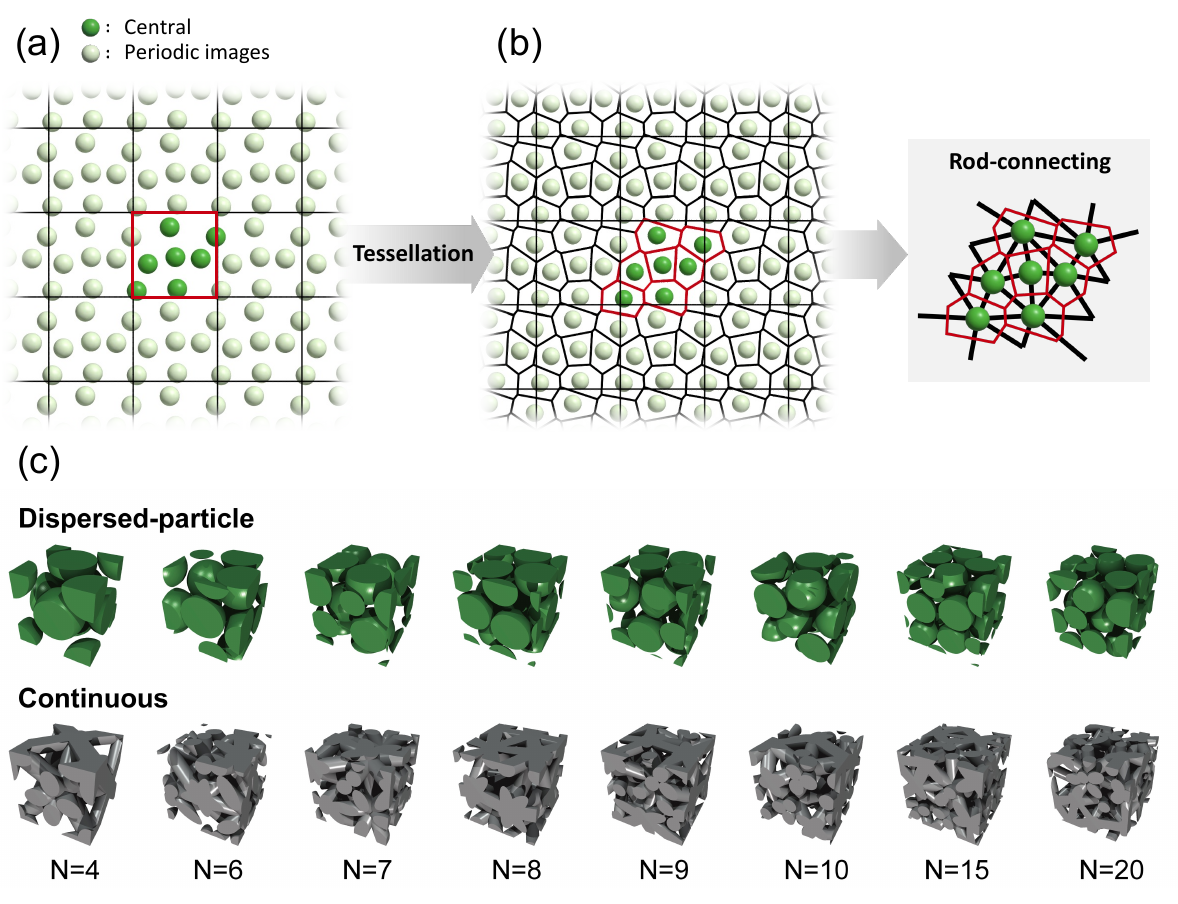}
    \caption{Constructing heterogeneous disordered materials with two distinct morphologies: (a) dispersed-particle morphology and (b) continuous morphology. (c) Representative volume elements (RVEs) with dispersed-particle and continuous morphologies with varying number of random spatial points (N = 4, 6, 7, 8, 9, 10, 15 and 20). Here, only hard domains are displayed in RVEs and note that v$_\mathrm{hard}$ = 50\%.}
    \label{fig:design}
\end{figure}

We summarize the design procedure used to create the heterogeneous disordered materials in Figure \ref{fig:design}. Figure \ref{fig:design}a schematically illustrates the two-dimensional analogy of a three-dimensional setting for constructing representative volume elements (RVEs) of heterogeneous disordered materials with a dispersed-particle morphology. The centers of N particles are placed randomly and sequentially within a unit cube. During the placement of the N particles, 26 periodic images are simultaneously created through offsetting by unity from the original coordinates. When placing a new particle into the unit cube, we check if it interferes with the existing particles and their periodic images; i.e., a trial coordinate for the new particle is accepted if it does not overlap with the existing particles or any of their periodic images. Once the centers of the N particles are determined inside the unit cube without any overlapping, a periodic RVE is constructed by taking the central unit cube from the supercell consisting of a 3 by 3 by 3 array of unit cells (e.g., highlighted by the red square in Figure \ref{fig:design}a). Next, to construct RVEs with continuous networks, we conduct Voronoi tessellation via the N randomly placed points under periodic boundary conditions, as shown in Figure \ref{fig:design}b; it should be noted that N-Voronoi tessellation is performed throughout the spatial points used to construct the RVEs with the dispersed-particle morphology. We then rod-connect the center points and their neighboring points (adjacent polyhedral cells sharing the tessellated planes) for all of the N points within the unit cube. The inset of Figure \ref{fig:design}b schematically depicts a Voronoi-tessellated central unit cell (red solid lines) with randomly connected network (black solid lines). When rod-connecting the center to the neighboring points, the cutoff radius for each polyhedral cell is taken to be 90\% of the maximum distance between the center point and the neighbors. Figure \ref{fig:design}c shows exemplar RVEs with the dispersed-particle morphology along with their continuous counterparts, each with varying numbers of random spatial points (N = 4, 6, 7, 8, 9, 10, 15 and 20). All RVEs used in this work are displayed in Figures \ref{fig:rve_disp} and~\ref{fig:rve_cont} in Appendix \ref{subsec:RVEs}. It should also be noted that the volume fraction of dispersed particles or continuous networks within each of the RVEs is taken to be 50\%; i.e., v$_\mathrm{hard}$ = 50\%. We then present combined results of experiments and numerical simulations on 3D-printed prototypes of the heterogeneous materials with these dispersed-particle and continuous RVEs. These prototypes are composed of “hard” particles or networks (thermoplastic polymer) and “soft” matrices (rubbery polymer); the hard components exhibit significant energy dissipation with high yield and flow stresses, while the soft components exhibit a relatively compliant stress-strain response with negligible energy dissipation. Details on the 3D-printed prototypes and constituent materials are provided in Appendices \ref{subsec:fabrication} and \ref{subsec:constitutive}.

\clearpage

\section{Elastic anisotropy: dispersed-particle vs continuous}
\label{sec:elasticAnisotropy}
First, we focus on elastic anisotropy in these heterogeneous disordered materials with varying numbers of random spatial points (N). Here, the volume fraction of the hard component was taken to be v$_\mathrm{hard}$ = 50\%; we note that the effective packing fraction, v$_\mathrm{pack}$ = 50.75\%, was used when placing the N spatial points for clearance between the hard particles\footnote{\label{note:packing}We also note that the simple sphere packing algorithm employed in this work was not able to construct a point configuration for N = 5 with this specific packing fraction; the maximum achievable packing fraction was found to be $\sim$ 47\% with N = 5 (see Figure~\ref{fig:q4q6_n2345}  in Appendix \ref{subsec:MaxPackingFraction}).}. The elastic anisotropy is examined for the RVEs with (1) the dispersed-particle morphology and (2) the corresponding continuous counterpart. We conducted a micromechanical analysis of the disordered RVEs subjected to periodic boundary conditions as well as macroscopic deformation conditions; detailed modeling procedures can be found in Appendix \ref{subsec:Micromechanical}. Figure \ref{fig:elastic_anisotropy} presents the elastic anisotropy of ten statistical realizations as a function of the random spatial points for both morphologies. The three-dimensional maps displayed in Figures \ref{fig:elastic_anisotropy}a and \ref{fig:elastic_anisotropy}c depict the directional (or effective) elastic modulus with respect to the crystallographic orientations in each of the RVEs with the two different morphologies; i.e., the more spherical the anisotropy map is, the more isotropic is the disordered material. In addition to the graphical representations of the variation in the magnitude of the directional elastic modulus, we computed the anisotropy indices: the universal anisotropy index (\cite{ranganathan2008universal}) and the equivalent Zener index (\cite{nye1985physical}) (see Appendix \ref{subsec:Micromechanical}). As shown in Figures \ref{fig:elastic_anisotropy}a and \ref{fig:elastic_anisotropy}b, the disordered RVEs with the dispersed-particle morphology exhibit strong elastic anisotropy with significant statistical variations, especially when N $<$ 10; furthermore, as expected, the elastic anisotropy is shown to decrease in the RVEs with the smaller stiffness ratio of the two constituents ($E_{\mathrm{soft}}/E_{\mathrm{hard}} = 1/30$). For N $\geq$ 10, the elastic anisotropy is rapidly reduced for both stiffness ratios ($E_{\mathrm{soft}}/E_{\mathrm{hard}} = 1/30$ and $1/90$). However, increasing the number of random spatial points in the dispersed-particle morphology is not shown to be an efficient means of altering the elastic anisotropy in these heterogeneous, disordered materials; i.e., $\mathrm{A}^{\mathrm{U}} = 0.041$ and $0.088$ ($\mathrm{A}^{\mathrm{eq}} = 1.20$ and $1.31$) for N = 20 with $E_{\mathrm{soft}}/E_{\mathrm{hard}} = 1/30$ and $1/90$, respectively, comparable to those in a dispersed-particle RVE constructed on a close-packed face-centered-cubic (fcc) lattice (as indicated by the dashed blue/orange lines in Figure \ref{fig:elastic_anisotropy}b). The significant elastic anisotropy observed across the statistical realizations of the dispersed-particle RVEs is mainly attributed to the small “finite” number of random points and the relatively high volume fraction of hard particles. Furthermore, with the strong geometric constraints involving periodic boundary conditions and the finite number of points (here, N $\leq$ 20), the volume fraction of the hard particles (v$_\mathrm{hard}$ = 50\% and v$_\mathrm{pack}$ = 50.75\%) is shown to be close to the jamming limit at which no additional particles can be added, leading to significant elastic anisotropy in the dispersed-particle RVEs. This is not very surprising as these RVEs do not attain the thermodynamic limit: the number of particles, N $\rightarrow \infty$, and the volume of systems, V $\rightarrow \infty$, necessary conditions for isotropy in jammed systems, which have been addressed in many previous studies of the dense packing of hard particles (\cite{torquato2000random, segurado2002numerical, torquato2010jammed, goodrich2012finite, goodrich2014jamming}).

Next, we examine the elastic anisotropy in the disordered RVEs with continuous hard domains. We further note that these continuous RVEs were derived from the same random point distributions used in their dispersed-particle counterparts. As shown in Figures \ref{fig:elastic_anisotropy}c and \ref{fig:elastic_anisotropy}d, adding connectivity throughout the randomly distributed spatial points results in a significant decrease in the elastic anisotropy in all N from 4 to 20. Furthermore, for both stiffness ratios ($E_{\mathrm{soft}}/E_{\mathrm{hard}} = 1/30$ and $1/90$), the statistical variations across the ten realizations are shown to diminish remarkably, compared to the corresponding ten dispersed-particle RVEs. Interestingly, the anisotropy maps and the anisotropy indices in the disordered RVEs with N = 4 are found to be very similar to those in the continuous “ordered” RVEs on fcc, also known as octet-truss lattice materials (\cite{deshpande2001effective, zheng2014ultralight, meza2014strong}). Moreover, the anisotropy map is shown to have a spinning-top shape with very low statistical variation for N = 6; this shape is nearly identical across the ten statistical realizations, and only rigid-body rotation without any shape change is observed. More importantly, an abrupt drop in the elastic anisotropy is observed between N = 6 and N = 7. The continuous, disordered RVEs exhibit near-complete elastic isotropy with N $\geq$ 7; $\mathrm{A}^{\mathrm{U}}$ = 0.012 $\pm$ 0.0094 (0.022 $\pm$ 0.015), 0.0280 $\pm$ 0.011 (0.047 $\pm$ 0.018), 0.037 $\pm$ 0.017 (0.065 $\pm$ 0.026), 0.022 $\pm$ 0.018 (0.039 ± 0.027), 0.0062 $\pm$ 0.0026 (0.010 ± 0.0037) and 0.0064 $\pm$ 0.0022 (0.0097 $\pm$ 0.0035) correspondingly for N = 7, 8, 9, 10, 15 and 20 with $E_{\mathrm{soft}}/E_{\mathrm{hard}} =$ 1/30 (1/90). The near-complete elastic isotropy in these RVEs with N $\geq$ 7 is shown to remain unaffected by the stiffness ratio of the constituent materials.

\begin{figure}[h!]
    \centering
    \includegraphics[width=1.0\textwidth]{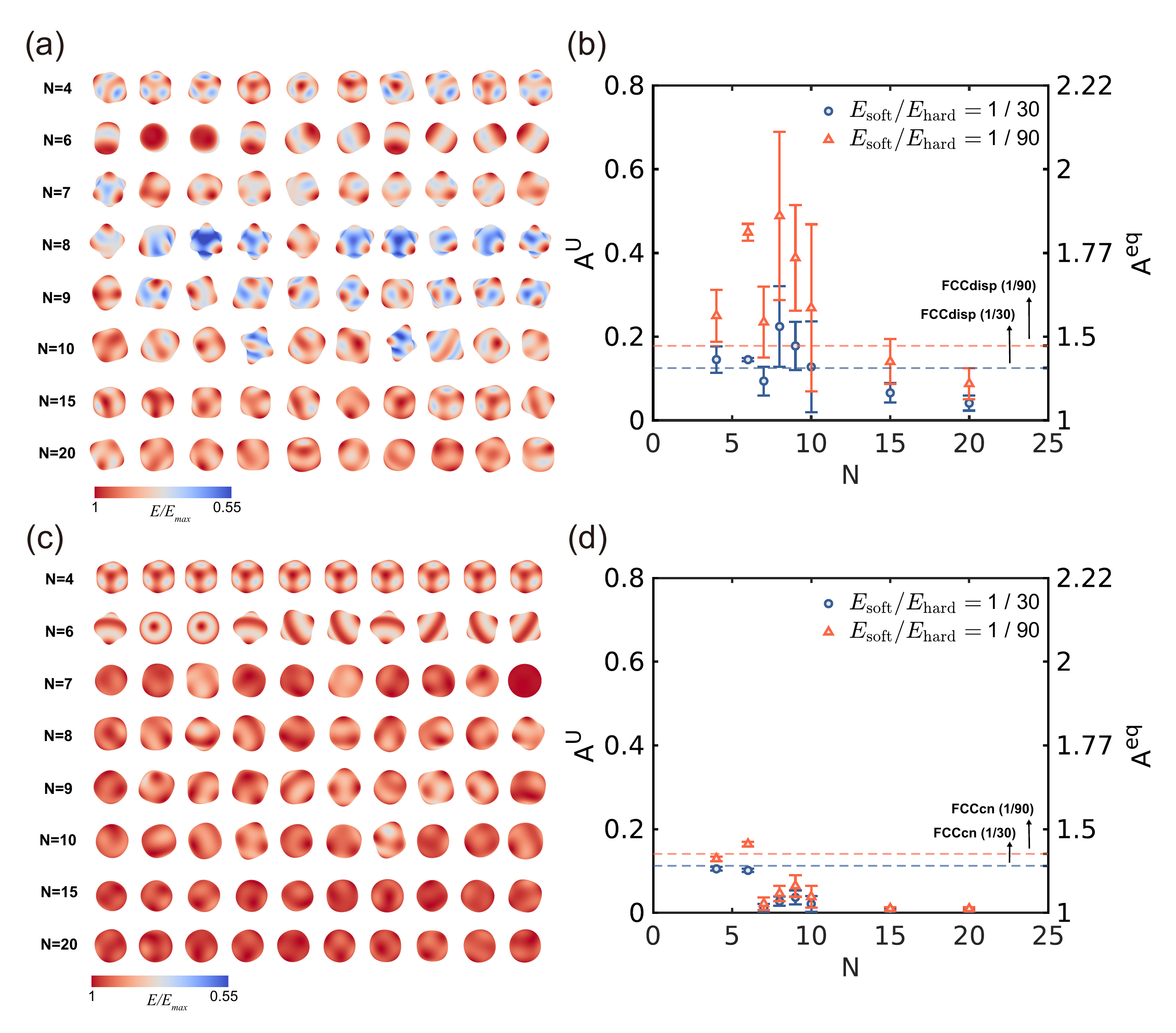}
    \caption{Elastic anisotropy in heterogeneous disordered materials with v$_\mathrm{hard}$ = 50\% (v$_\mathrm{pack}$ = 50.75\%)\protect\textsuperscript{\ref{note:packing}}. (a) Three-dimensional maps of the directional elastic modulus with $E_{\mathrm{soft}}/E_{\mathrm{hard}} = 1/30$ and (b) anisotropy indices in RVEs with dispersed hard domains with two different stiffness ratios ($E_{\mathrm{soft}}/E_{\mathrm{hard}}$) of 1/30 and 1/90. Here, $\mathrm{A}^{\mathrm{U}}$ is the universal anisotropy index and $\mathrm{A}^{\mathrm{eq}}$ is the equivalent Zener index (\cite{ranganathan2008universal,nye1985physical}). The blue and orange dashed lines indicate the anisotropy indices in the dispersed-particle RVEs constructed on a face-centered-cubic (fcc) lattice. (c) Three-dimensional maps of the directional elastic modulus with $E_{\mathrm{soft}}/E_{\mathrm{hard}} = 1/30$ and (d) anisotropy indices in the RVEs with continuous hard domains with $E_{\mathrm{soft}}/E_{\mathrm{hard}} = 1/30$ and $1/90$. The blue and orange dashed lines indicate the anisotropy indices in the continuous RVEs on a fcc lattice.}
    \label{fig:elastic_anisotropy}
\end{figure}
\clearpage

\section{Microstructural analysis: roles of connectivity, rod length and local symmetry}
\label{sec:Microstructural}
\begin{figure}[h!]
    \centering
    \includegraphics[width=1.0\textwidth]{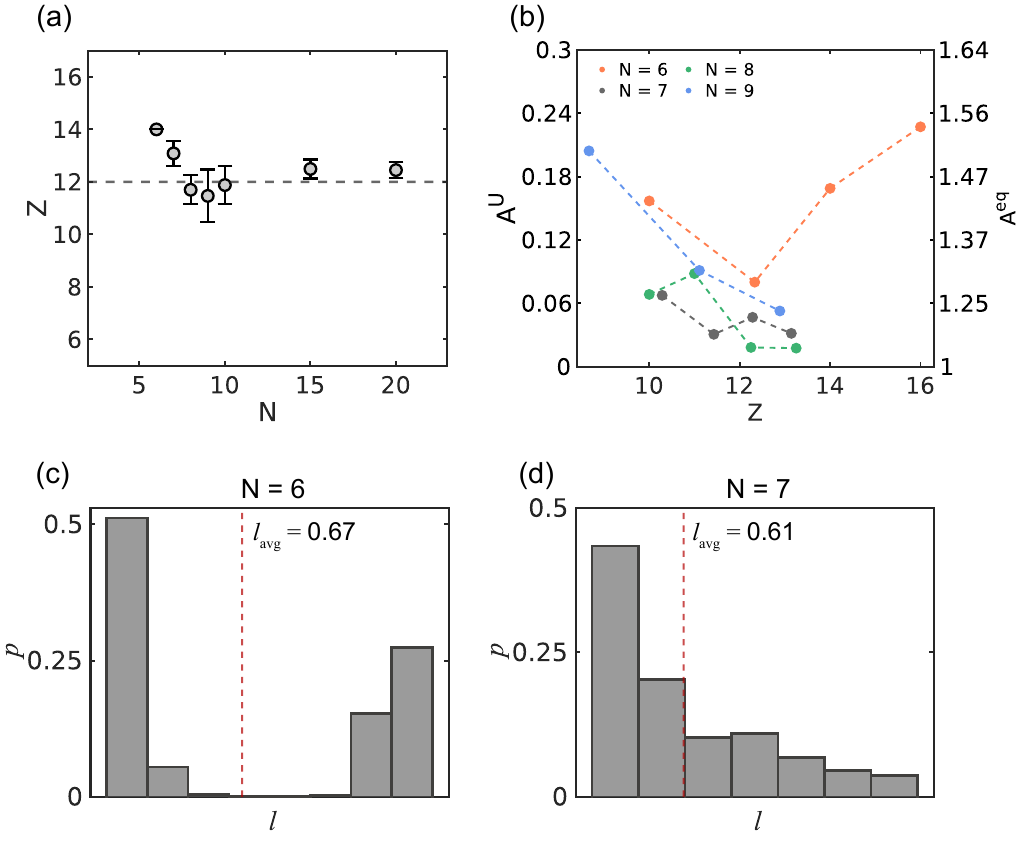}
    \vspace{-0.3in}
    \caption{Microstructural details in heterogeneous disordered materials with continuous morphology. (a) Average number of connectivities (or the coordination number, Z) as a function of the number of random spatial points; the dashed line indicates the coordination number in a fcc lattice (Z = 12). (b) Elastic anisotropy of the RVEs with continuous morphology with varying coordination number. Here, note that a RVE with the highest anisotropy was selected for analysis from the ten statistical realizations for each of N = 6, 7, 8 and 9. Rod length distributions in the continuous RVEs with (c) N = 6 and (d) N = 7, obtained from ten statistical realizations which are displayed in Figure \ref{fig:design}c; here, the red dashed line indicates the average rod length in each of N = 6 and N = 7. Note that v$_\mathrm{hard}$ = 50\% (v$_\mathrm{pack}$ = 50.75\%).}
    \label{fig:rods}
\end{figure}

In Figure \ref{fig:elastic_anisotropy}, we demonstrated that altering the connectivity throughout the random spatial points significantly influences the elastic anisotropy in these heterogeneous disordered materials. The “continuous” RVEs show remarkably low elastic anisotropy with N $\geq$ 7. Microstructural details underlying the transition between N = 6 and N = 7 in the disordered materials with the continuous hard morphology as well as with equal volume fractions of the constituents (v$_\mathrm{hard}$ = 50\% but v$_\mathrm{pack}$ = 50.75\% due to clearance) are displayed in Figure \ref{fig:rods}. Figure \ref{fig:rods}a shows the average number of connectivities (or the coordination number, Z; \cite{muller1994glossary}) obtained from ten statistical realizations of each of the RVEs with varying numbers of random spatial points; the number of connectivities has been found to play a crucial role in elastic and inelastic features in a broad variety of structured materials with long-range ordering at small (\cite{deshpande2001foam, fleck2010micro, greer2019three}) to large strains (\cite{wang2011co, cho2016engineering, lee2024extreme}). Interestingly, the average coordination number in all RVEs with N = 6, 7, 8, 9, 10, 15 and 20 is found to be Z $\sim$ 12, close to that in octet-truss materials on a face-centered-cubic (fcc) lattice, thus revealing that the underlying microstructures derived from the random sphere packing are nearly close-packed. However, the coordination number is not shown to provide sufficient geometric information to account for the sudden drop in the elastic anisotropy we observed between N = 6 and 7. Further, as demonstrated in Figure \ref{fig:rods}b, increasing the coordination number (Z $>$ 12) by increasing the cutoff radius for rod-connecting is not an efficient means of achieving elastic isotropy ($\mathrm{A}^{\mathrm{U}}$ $\rightarrow$ 0); rather, across the RVEs with varying numbers of random spatial points, Z $\sim$ 12  is shown to be a nearly optimal coordination number for elastic isotropy.

Next, the histograms plotted in Figures \ref{fig:rods}c and \ref{fig:rods}d show the distributions of lengths of rods in the continuous RVEs with N = 6 and N = 7, analogous to a pair correlation function (or a radial distribution function) that describes the number density variation as a function of the distance from each of the random spatial points. The distribution of rod lengths in the RVEs with N = 6 (Figure \ref{fig:rods}c) is shown to be bimodal, attributed to the relatively sparse random points; points far from each of the Voronoi center points are more likely to be identified as neighbors. Long rods connecting distant neighbors are shown to give rise to the high anisotropy observed in the RVEs with N = 6 ($\mathrm{A}^{\mathrm{U}}$ = 0.10 $\pm$ 0.0037 and 0.16 $\pm$ 0.0058 for $E_{\mathrm{soft}}/E_{\mathrm{hard}} = 1/30$ and $1/90$, respectively). However, as shown in Figure \ref{fig:rods}d, the distribution obtained from the ten statistical realizations with N = 7 is shown to be positive-skewed with a relatively long, right tail. This type of positively skewed distribution is observed across all of the statistical realizations with N = 7, 8, 9, 10, 15 and 20, as displayed in Figure~\ref{fig:rod_distributions} in Appendix \ref{subsec:rod_length}, analogous to the radial distribution function widely observed in statistically isotropic many-body systems (\cite{yuste1991radial, rintoul1997reconstruction, bhattacharjee2003nth, morsali2005accurate, terban2021structural}). When N $\geq$ 7, densely packed, here with v$_\mathrm{hard}$ = 50\% yet v$_\mathrm{pack}$ = 50.75\%, the hard network comprises relatively short rods throughout the random spatial points, resulting in very low elastic anisotropy.

\clearpage

\begin{figure}[h!]
    \centering
    \includegraphics[width=1.0\textwidth]{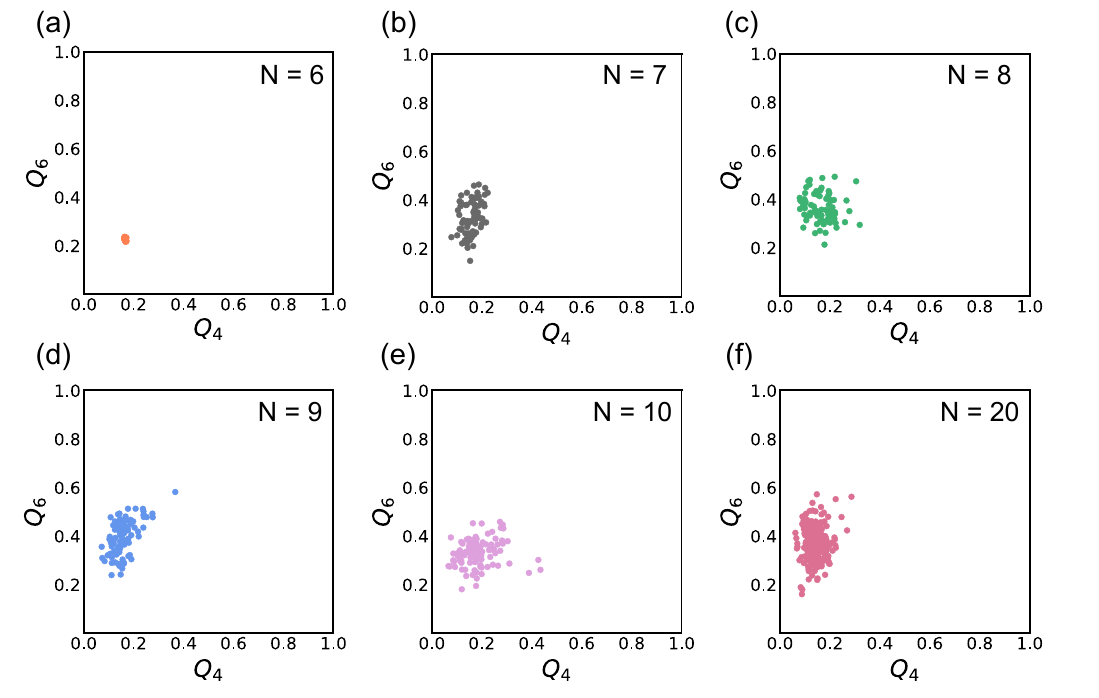}
    \caption{Local bond-orientational order parameters ($Q_4$, $Q_6$) in ten statistical realizations with (a) N = 6, (b) N = 7, (c) N = 8, (d) N = 9, (e) N = 10 and (f) N = 20. Note that v$_\mathrm{hard}$ = 50\% (v$_\mathrm{pack}$ = 50.75\%).}
    \label{fig:q4q6}
\end{figure}

The microstructural details in these heterogeneous materials are further explored in Figures \ref{fig:q4q6}a--\ref{fig:q4q6}f, presenting the bond-orientational orders, $Q_4$ and $Q_6$, in the statistical realizations with N = 6, 7, 8, 9, 10 and 20. These indices are widely used for quantifying the degree of crystallization in condensed matter (\cite{steinhardt1983bond, leocmach2012roles, tanaka2019revealing, tang2021synthesis}). 
The bond-orientational order parameter $Q_l$ of a symmetry order $l \geq 0$ is given by, $Q_l=\sqrt{\frac{4 \pi}{2 l+1} \sum_{m=-l}^l\left|\left\langle Y_{l m}\left(\theta\left(\stackrel{\rightharpoonup}{r}_i\right), \phi\left(\stackrel{\rightharpoonup}{r}_i\right)\right)\right\rangle\right|^2}$ 
where $Y_{lm}$ is the spherical harmonics for the bond $\stackrel{\rightharpoonup}{r}$ between two points, $\theta(\stackrel{\rightharpoonup}{r})$ is the polar angle of $\stackrel{\rightharpoonup}{r}$, $\phi(\stackrel{\rightharpoonup}{r})$ is the polar angle of $\stackrel{\rightharpoonup}{r}$ and $-l \leq m \leq l$ is the orientation with respect to a reference set of orthonormal axes. Also, $\left\langle \cdots \right\rangle$ denotes an average over all bonds $\stackrel{\rightharpoonup}{r}_i$. Note that $Q_l$, which represents the strength of $l$-fold symmetry, is invariant under rotations of coordinate systems. In this study, we used $Q_4$ and $Q_6$ to characterize the local orientational symmetries in our disordered RVEs. Convergence in these bond orientational orders indicates that the network in these RVEs has local orientational symmetry; e.g., in fcc crystals, ($Q_4$, $Q_6$) = (0.191, 0.574). Interestingly, ($Q_4$, $Q_6$) in N = 6 (Figure \ref{fig:q4q6}a) is shown to converge on (0.165, 0.225), thus revealing that the underlying microstructures are crystallized at this specific packing fraction (v$_\mathrm{pack}$ = 50.75\%) despite the fact that they were derived from random sphere packing. This is further evidenced in Figure \ref{fig:elastic_anisotropy}c (or Figure \ref{fig:elastic_anisotropy}a on the dispersed-particle RVEs), where the spinning-top shaped anisotropy map is nearly identical across the ten statistical realizations. In contrast, for N $\geq$ 7, the bond orientational orders are found to disperse widely in $0.1 < Q_4 < 0.3$ and $0.2 < Q_6 < 0.5$ (Figures \ref{fig:q4q6}b--\ref{fig:q4q6}f), thus revealing that the underlying microstructures have no significant orientational symmetry at this point. 

\begin{figure}[h]
    \centering
    \includegraphics[width=0.6\textwidth]{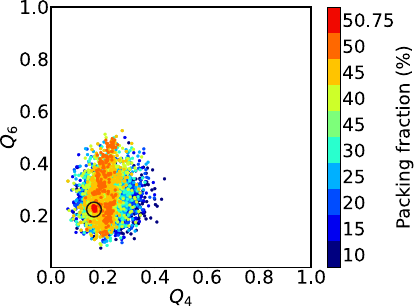}
    \linespread{1.0}
    \caption{Local bond-orientational order parameters ($Q_4$, $Q_6$) in statistical realizations with N = 6 with varying packing fraction from v$_\mathrm{pack}$ = 10\% to 50.75\%}
    \label{fig:q4q6_n6}
\end{figure}

\begin{figure}[h!]
    \centering
    \includegraphics[width=1.0\textwidth]{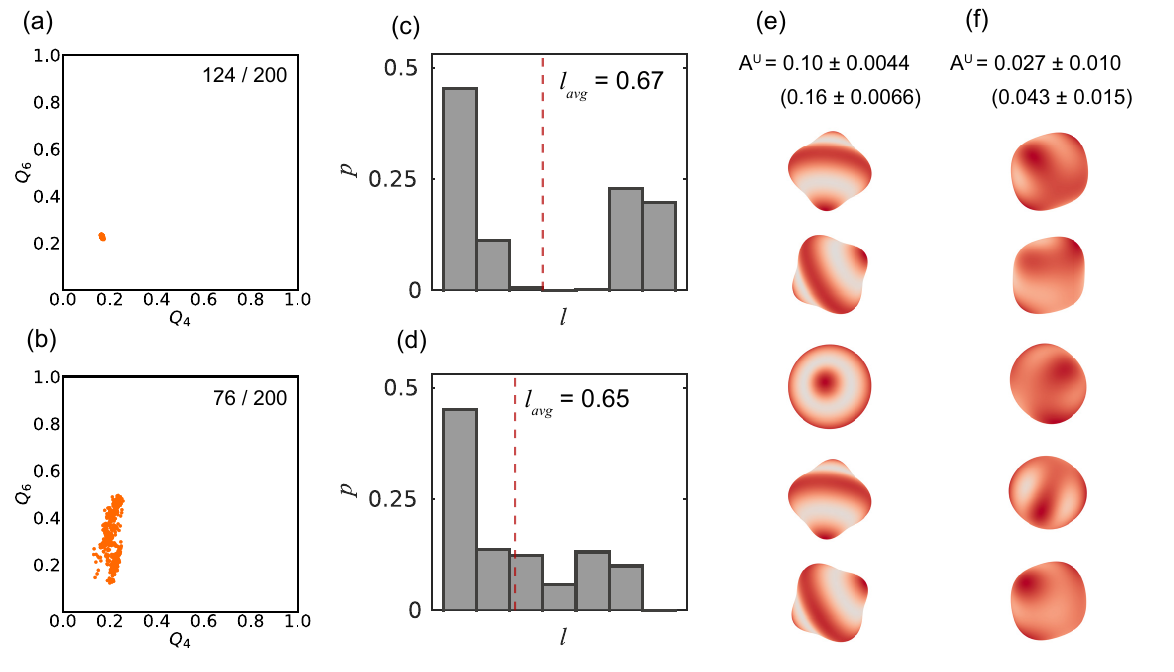}
    \linespread{1.0}
    \vspace{-0.3in}
    \caption{Microstructural details in the continuous RVEs (200 statistical realizations) with N = 6 at v$_\mathrm{pack}$ = 50\%. Local bond-orientational orders in (a) 124 statistical realizations that exhibit significant orientational symmetry and (b) 76 statistical realizations which are not crystallized. The rod length distributions in (c) crystallized 124 RVEs and (d) non-crystallized 76 RVEs.  The directional elastic moduli in five representative realizations (e) crystallized and (f) non-crystallized with $E_{\mathrm{soft}}/E_{\mathrm{hard}}$ = 1/30 (1/90).}
    \label{fig:n6_details}
\end{figure}

As evidenced in Figures \ref{fig:elastic_anisotropy} and \ref{fig:q4q6}, the continuous RVEs with N = 6 are shown to have local orientational symmetry at v$_\mathrm{pack}$ = 50.75\%. Figure \ref{fig:q4q6_n6} shows the bond-orientational orders in the point configurations (again from one hundred statistical realizations) for N = 6 with varying packing fractions from 10\% to 50.75\%. At relatively lower packing fractions, the bond-orientational orders are found to disperse widely, indicating that the microstructures in these RVEs have no significant orientational symmetry. However, ($Q_4$, $Q_6$) begins to converge at v$_\mathrm{pack}$ = 50\%, revealing that the point configurations for N = 6 crystallize at this particular packing fraction, which identical to the volume fraction of hard components investigated throughout this work. Furthermore, at v$_\mathrm{pack}$ = 50.75\%, a nearly maximum achievable packing fraction for N = 6, all statistical realizations exhibit orientational symmetry with ($Q_4$, $Q_6$) = (0.165, 0.225), i.e., they are crystallized. We investigated the microstructures within these RVEs with N = 6 further by constructing new, continuous RVEs with v$_\mathrm{pack}$ = 50\% without any clearance (v$_\mathrm{hard}$ = 50\%). As shown in Figures \ref{fig:n6_details}a and \ref{fig:n6_details}b, 124 statistical realizations out of 200 exhibit significant orientational symmetry with ($Q_4$, $Q_6$) $\sim$ (0.165, 0.225), while the remaining 76 statistical realizations are not crystallized. Accordingly, strong elastic anisotropy ($\mathrm{A}^{\mathrm{U}}$ = 0.10 $\pm$ 0.0044 and 0.16 $\pm$ 0.0066 for $E_{\mathrm{soft}}/E_{\mathrm{hard}} = 1/30$ and $1/90$, respectively) with spinning-top-shaped directional stiffness is observed across five representative crystallized RVEs (from 124 realizations, Figure \ref{fig:n6_details}e). Elastic anisotropy significantly decreases in the five representative RVEs (from 76 realizations) that exhibit dispersed $Q_4$ and $Q_6$ values, as displayed in Figure \ref{fig:n6_details}f. Additionally, the distributions of rod lengths obtained from the 124 statistical realizations in Figure \ref{fig:n6_details}c are found to be bimodal, further supporting the significant elastic anisotropy by virtue of the crystallized microstructures. However, the microstructures in the 76 remaining non-crystallized statistical realizations are shown to have positive-skewed distributions (Figure \ref{fig:n6_details}d), giving rise to much lower elastic anisotropy in these RVEs with the same packing fraction, v$_\mathrm{pack}$ = 50\% (v$_\mathrm{hard}$ = 50\%). The small change in the packing fraction (v$_\mathrm{pack}$ = 50\% $\rightarrow$ v$_\mathrm{pack}$ = 50.75\%) is found significantly to impact the elastic anisotropy highly sensitive to geometric features characterized by the rod length distributions and the bond-orientation orders in the RVEs with N = 6, especially near the corresponding maximum achievable packing fraction, which is approximately 50.75\%. 

\vspace{0.2in}

\clearpage

\section{Large strain elastic and inelastic behavior}
\label{sec:LargeStrain}
\begin{figure}[h!]
    \centering
    \includegraphics[width=1.0\textwidth]{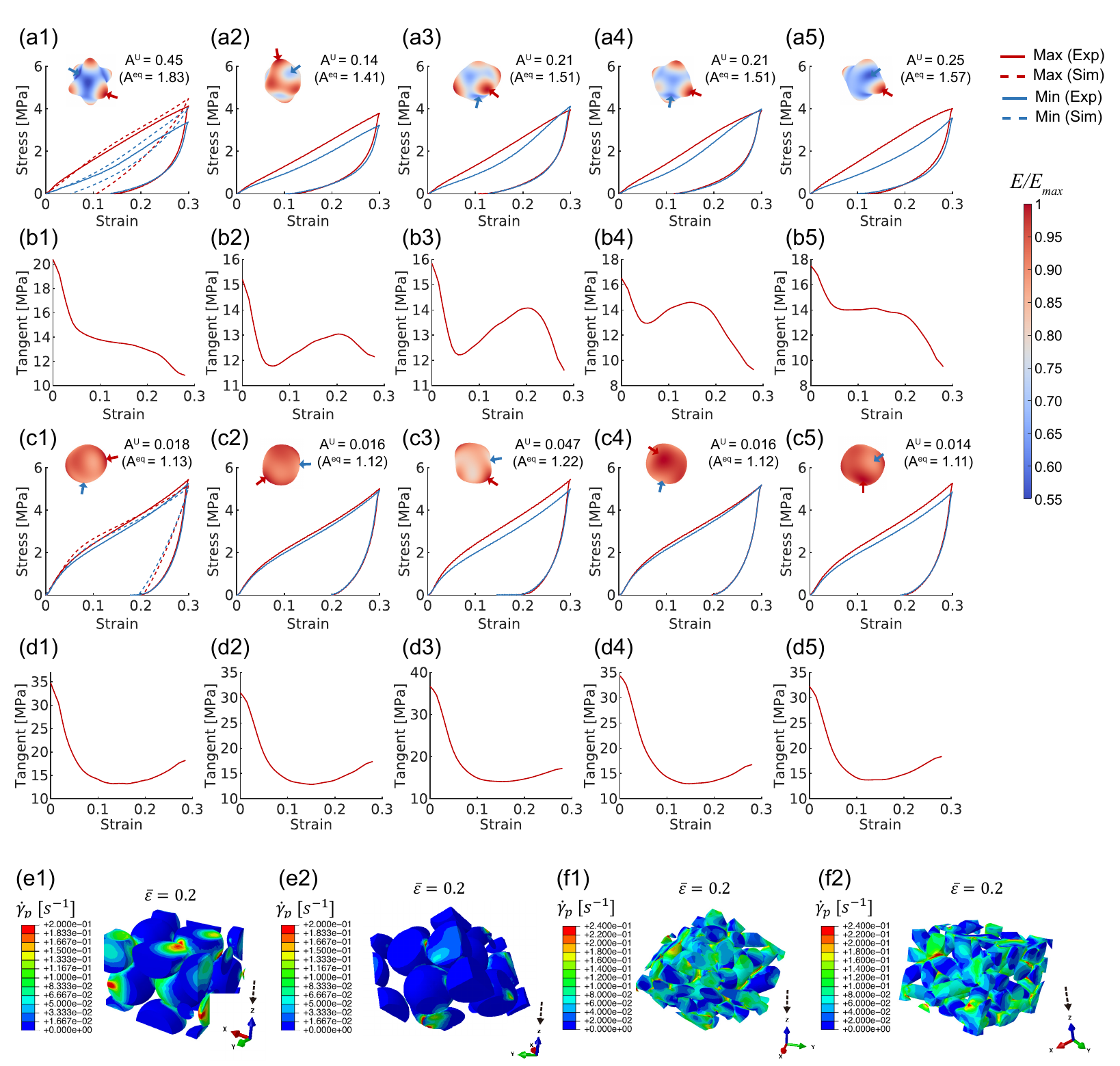}
    \caption{Effect of adding connectivity on the large strain mechanical behaviors in heterogeneous disordered materials. (a1-a5) Stress-strain curves in five representative 3D-printed prototypes with dispersed-particle morphology, loaded in two directions of the maximum and minimum elastic moduli (solid lines: experiment, dashed lines: numerical simulation; insets: corresponding directional elastic modulus map). (b1-b5) Instantaneous tangent modulus in five representative 3D-printed prototypes loaded in the maximum modulus direction. (c1-c5) Stress-strain curves in five representative 3D-printed prototypes  with continuous morphology, loaded in two directions of the maximum and minimum elastic moduli (solid lines: experiment, dashed lines: numerical simulation; insets: corresponding directional elastic modulus map). (d1-d5) Instantaneous tangent modulus in five representative 3D-printed prototypes in the maximum modulus direction. Contours of plastic strain rates in hard domains: a dispersed-particle RVE loaded in (e1) maximum and (e2) minimum modulus directions, and a continuous RVE loaded in (f1) maximum and (f2) minimum modulus directions at an imposed macroscopic strain of 0.2.}
    \label{fig:large_deformation}
\end{figure}

The connectivity throughout the hard domains significantly impacted the elastic isotropy of heterogeneous disordered materials; importantly, we numerically demonstrated that near-complete elastic isotropy can be achieved with N $\geq$ 7 in these materials with equal fractions of hard and soft components through adding the connectivity. Here, we probe the isotropy in these materials further under large strains beyond the infinitesimal elastic regime in both experiments and numerical simulations. More specifically, we investigate the highly resilient yet dissipative large strain features in these materials subjected to cyclic loading conditions. Towards this end, we fabricated prototypes of five representative statistical realizations for two different morphologies (dispersed-particle vs. continuous hard domains; number of random spatial points, N = 7) using a high-precision, multi-material 3D printer (Connex3 Objet260, Stratasys Inc.). A thermoplastic polymer was used for the hard domains while a rubbery polymer was used for the soft domains. The “initial” stiffness ratio of the constituent materials was $E_{\mathrm{soft}}/E_{\mathrm{hard}} = 1/90$. Detailed information about the 3D printing and experimental procedures can be found in Appendix \ref{subsec:fabrication}. Note that the hard components exhibit elastic-plastic stress-strain behavior with an apparent yield and high flow stresses while the soft components are shown to be hyperelastic with no significant energy dissipation. Moreover, highly nonlinear stress-strain behavior in each of the constituent hard and soft materials is presented together with the large strain constitutive modeling procedure used in our numerical simulations in Appendix \ref{subsec:constitutive} (Figures~\ref{fig:materials_exp},~\ref{fig:materials_model} and Table~\ref{Tab:material_parameter}). Figure \ref{fig:large_deformation} presents the large strain mechanical behaviors in these prototypes with dispersed-particle and continuous hard domains loaded in two directions of the maximum and minimum “initial” elastic moduli, as determined from the anisotropy map (insets of Figures \ref{fig:large_deformation}a and \ref{fig:large_deformation}c). During loading, the prototypes initially exhibit an elastic response, followed by yield-like stress-rollover and strain hardening; they also display highly nonlinear unloading behavior that involves a large amount of energy dissipation as well as a significant residual strain. As shown in Figure \ref{fig:large_deformation}a on a dispersed-particle morphology, the statistical realizations loaded in the maximum modulus directions exhibit much greater stress responses at increasing strains, compared to those in the minimum modulus directions, revealing significant anisotropy at small to large strains. Further, energy dissipation mainly due to plastic flows throughout the hard particles is found to be greater in the statistical realizations loaded in their maximum modulus directions. This is further confirmed by the plastic strain rate contours throughout the hard particles plotted in Figures \ref{fig:large_deformation}e1 and \ref{fig:large_deformation}e2. Plastic flows with much greater magnitudes (or flow strengths) are shown to be localized throughout adjacent hard particles in the RVE loaded in the corresponding maximum elastic modulus direction (Figure \ref{fig:large_deformation}e1). However, the hard particles in the RVE loaded in their corresponding minimum elastic modulus direction (Figure \ref{fig:large_deformation}e2) exhibit very weak plastic flows. The highly anisotropic elastic and inelastic features in these RVEs with the dispersed-particle morphology at small to large strains mainly result from the non-uniform anisotropic distribution of “finite” hard particles (here, N = 7); i.e., when the hard particles are well aligned with the loading direction, the materials exhibit relatively stiff (or plastically hard) responses. However, if the hard particles are staggered rather than stacked, they exhibit compliant (or plastically soft) behaviors, as clearly indicated in the measured and numerically simulated stress-strain curves in the statistical realizations loaded in their minimum elastic modulus directions. The stress-strain data are reduced further into an instantaneous tangent modulus at an increasing level of strain ($\frac{\partial \sigma}{\partial \varepsilon}$, where $\sigma$ is the axial stress and $\varepsilon$ is the axial strain) in Figure \ref{fig:large_deformation}b, especially for the five statistical realizations loaded in the maximum modulus directions. The tangent modulus is found to drop suddenly at imposed strains of $\sim$ 0.02 and $\sim$ 0.2; the first drop is mainly due to the “initial” plastic yield in the hard particles. The subsequent second drop in the tangent modulus is shown to be attributed to buckling instability throughout the hard particles well aligned with the loading direction, as extensively explored in previous studies, especially those focusing on particulate heterogeneous materials (\cite{tordesillas2009modeling, li2019domain, chen2023post, arora2024magnetically}).

Next, we examine the large strain behavior in heterogeneous disordered materials with continuous hard domains. As shown in Figure \ref{fig:large_deformation}c, adding connectivity throughout the random spatial points is shown to result in highly isotropic elastic and inelastic responses during loading and unloading; the stress-strain curves in the minimum modulus directions are very similar to those in the maximum modulus directions across all statistical realizations. The near-complete isotropic elastic and inelastic features involving the initial elastic modulus, inelastic yield, flow stress, energy dissipation and residual strain upon unloading are further evidenced in Figure \ref{fig:inelastic} in Appendix \ref{sec:inelastic_isotropy}. The effect of adding connectivity throughout the random spatial points on the isotropic inelastic features is also shown in Figures \ref{fig:large_deformation}f1 and \ref{fig:large_deformation}f2. In both the maximum and minimum modulus directions, plastic flows are found to develop uniformly throughout the hard ligament networks; this is quite different from what we have observed in the dispersed-particle morphology (Figures \ref{fig:large_deformation}e1 and \ref{fig:large_deformation}e2), where plastic flows with much greater strength were found to be localized throughout the hard particles aligned in the maximum modulus direction. Additionally, the stress-strain curves (Figure \ref{fig:large_deformation}c) and the corresponding tangent moduli (Figure \ref{fig:large_deformation}d) demonstrate no buckling instability in these prototypes with continuous hard networks; i.e., no “second” drop in the instantaneous tangent modulus is observed in any of the representative statistical realizations loaded in their maximum modulus directions. By adding connectivity, highly isotropic elastic and inelastic features with outstanding stability have been achieved in these heterogeneous disordered materials at small to large strains and during loading and unloading conditions.

\clearpage

\section{Cyclic behavior and reusability}
\label{sec:Cyclic}
\begin{figure}[h!]
    \vspace{-0.5in}
    \centering
    \includegraphics[width=0.95\textwidth]{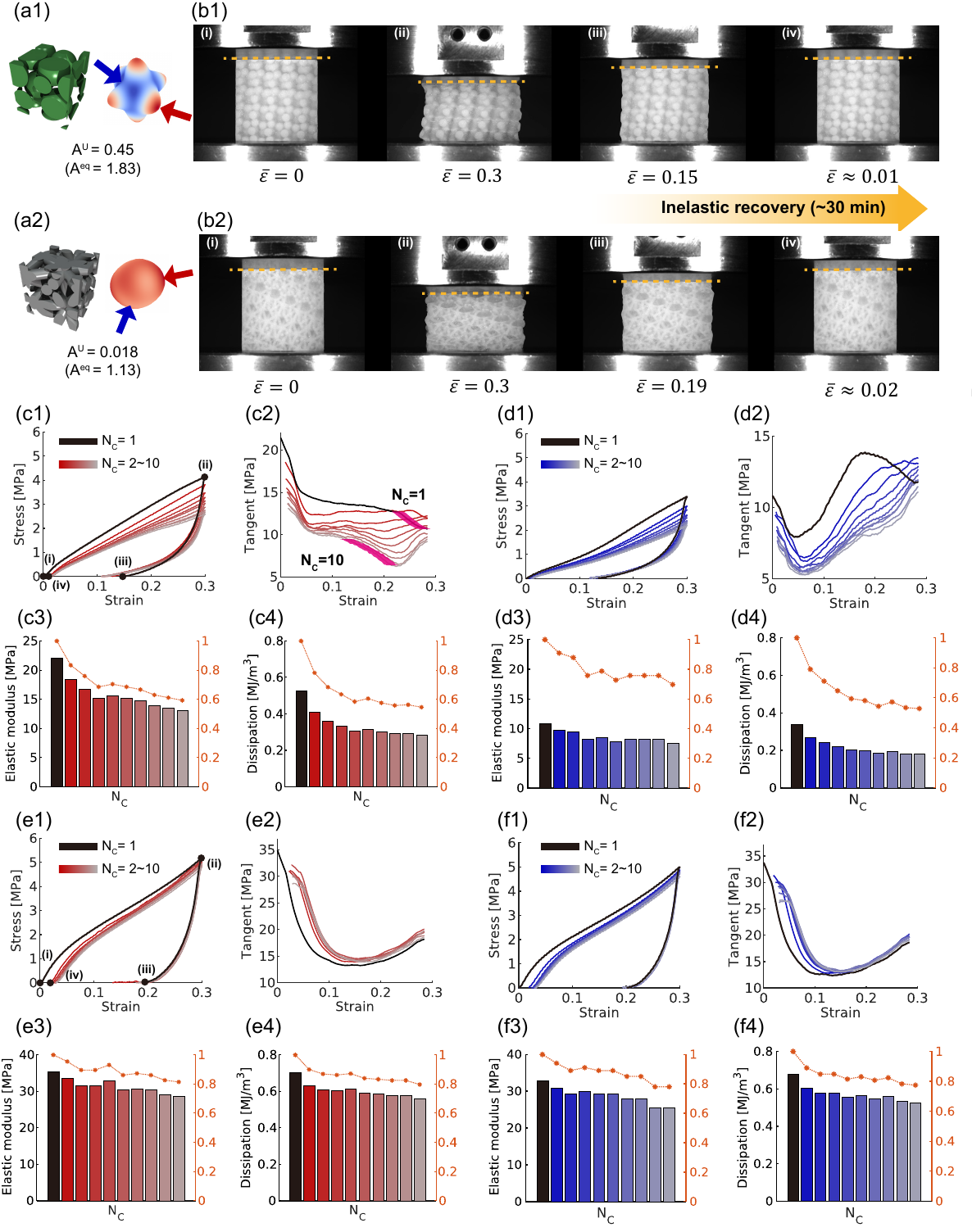}
    \vspace{-0.2in}
    \caption{Reusability of heterogeneous disordered materials during multiple loading and unloading cycles. Representative volume elements (again, only hard domains shown) and corresponding anisotropy maps for (a1) dispersed-particle and (a2) continuous morphologies. Experimental images of 3D-printed prototypes with (b1) dispersed-particle and (b2) continuous morphologies during the first cycle (N$_\mathrm{C}$ = 1): (i) at undeformed, (ii) at the maximum imposed strain ($\varepsilon$ = 0.3), (iii) at the end of elastic unloading and (iv) after idling time of 30 minutes for shape recovery. Stress-strain curves in a representative prototype with dispersed-particle hard domains loaded in its (c1) maximum and (d1) minimum modulus directions, together with the corresponding (c2) and (d2) instantaneous tangent moduli, (c3) and (d3) elastic moduli, (c4) and (d4) energy dissipation densities, all from N$_\mathrm{C}$ = 1 to N$_\mathrm{C}$ = 10. Stress-strain curves in a representative prototype with continuous hard domains loaded in its (e1) maximum and (f1) minimum modulus directions, together with the corresponding (e2) and (f2) instantaneous tangent moduli, (e3) and (f3) elastic moduli, (e4) and (f4) energy dissipation densities, all from N$_\mathrm{C}$ = 1 to N$_\mathrm{C}$ = 10.}
    \label{fig:cyclic}
\end{figure}

Subsequently, we examined resilience and dissipation in heterogeneous disordered materials subjected to multiple loading and unloading cycles. A representative volume element that exhibits the highest elastic anisotropy was selected from the ten statistical realizations for the multiple cyclic mechanical test; the RVEs and corresponding anisotropy maps for both dispersed-particle and continuous morphologies are displayed in Figure \ref{fig:cyclic}a. Figure \ref{fig:cyclic}b displays sequential images of two 3D-printed prototypes loaded in their maximum modulus directions: loading (i $\rightarrow$ ii), unloading (ii $\rightarrow$ iii; or elastic recovery) and idling for inelastic recovery (iii $\rightarrow$ iv), especially during the first cycle (N$_\mathrm{C}$ = 1). As shown in Figures \ref{fig:large_deformation}a, \ref{fig:large_deformation}c and \ref{fig:cyclic}b, the prototypes with the continuous morphology exhibit increased residual strains relative to those with the dispersed-particle morphology at the end of unloading, mainly due to the more pronounced plastic flows throughout the hard ligament network; i.e., plasticity better developed within the continuous morphology is shown to increase the energy dissipation whereas it negatively affects the elastic shape recovery ability (ii $\rightarrow$ iii). However, during the idling time (here, 30 minutes), both prototypes are found mostly to recover their original shape without any further physical treatment. Such “inelastically” driven shape recovery between the cycles has been extensively reported in many two-phase elastomeric materials consisting of both hard and soft domains (\cite{deschanel2009rate, cho2013constitutive, cho2017deformation, lee2023polyurethane, cho2024large}). Figure \ref{fig:cyclic}c1 presents the stress-strain curves of a 3D-printed prototype with dispersed-particle hard domains in multiple consecutive cycles (N$_\mathrm{C}$ = 1 to N$_\mathrm{C}$ = 10); note that the prototype was loaded in its maximum modulus direction and that the idling time was set to 30 minutes between the cycles. The experimental image at an imposed strain of 0.3 displayed in Figure \ref{fig:cyclic}b1 clearly shows macroscopic buckling throughout the stacked hard particles; instability is further evidenced in the instantaneous tangent modulus plotted in Figure \ref{fig:cyclic}c2 (black solid line, N$_\mathrm{C}$ = 1), where a second drop at an imposed strain of 0.2 is observed. We also noted that buckling throughout the hard particles in this particular maximum modulus loading direction becomes more pronounced in the subsequent cycles from N$_\mathrm{C}$ = 2 to N$_\mathrm{C}$ = 10 (see also Movie S1 in Appendix \ref{sec:Movies}). As highlighted in Figure \ref{fig:cyclic}c2, the critical strain for instability is found to decrease from $\sim$ 0.2 (N$_\mathrm{C}$ = 1) to $\sim$ 0.15 (N$_\mathrm{C}$ = 10). Further, the load transfer and energy dissipation capabilities in the prototype are shown to degrade significantly in the subsequent cycles of N$_\mathrm{C}$ = 2 to 10, as further revealed by the elastic modulus ($E$) and the dissipated work density ($D$) as functions of N$_\mathrm{C}$ displayed in Figures \ref{fig:cyclic}c3 and \ref{fig:cyclic}c4, where $E_{10}/E_1$ = 0.59 and $D_{10}/D_1$ = 0.55. Additionally, as shown in Figure \ref{fig:cyclic}d1, the prototype loaded in the minimum modulus direction exhibits much smaller stress responses than in the maximum modulus direction during multiple cycles. Interestingly, a second drop in the tangent modulus is observed near an imposed strain of 0.2 as well in this minimum modulus direction, especially during the first cycle (black solid line in Figure \ref{fig:cyclic}d2), presumably due to bucking instability after densification of the staggered hard particles. Instability in the minimum modulus direction is shown to be mitigated in the subsequent cycles from N$_\mathrm{C}$ = 2 to N$_\mathrm{C}$ = 10 (see Movie S2 in Appendix \ref{sec:Movies}). Though the prototype is shown to be stable with no buckling (or second drop in the tangent modulus) during the subsequent cycles, it undergoes gradual degradation in load transfer and energy dissipation capabilities, as plotted in Figures \ref{fig:cyclic}d3 and \ref{fig:cyclic}d4, where $E_{10}/E_1$ = 0.69 and $D_{10}/D_1$ = 0.53.

Last, we turn our attention to the reusability of the prototype with continuous hard domains that was found to exhibit highly isotropic energy dissipation and load transfer capabilities (Figure \ref{fig:large_deformation}). Figures \ref{fig:cyclic}e and \ref{fig:cyclic}f present the cyclic behaviors of the prototype loaded in its maximum and minimum modulus directions, respectively. The stress-strain curves in the direction of the minimum elastic modulus are nearly identical to those in the maximum elastic modulus, supporting the near-complete isotropic resilience and dissipation during repeated cycles (Figures \ref{fig:cyclic}e1 and \ref{fig:cyclic}f1). Additionally, there is no instability throughout the hard ligament network in these prototypes loaded in both the maximum and minimum modulus directions during multiple cycles. Highly stable deformation is also clearly displayed in the sequential images in Figure \ref{fig:cyclic}b2 (see Movies S3 and S4 in Appendix \ref{sec:Movies}). The connectivity throughout the neighboring points enables the suppression of buckling instability in any loading direction, as further evidenced by the tangent moduli in multiple cycles in Figures \ref{fig:cyclic}e2 and \ref{fig:cyclic}f2. In addition to high stability in both loading directions, the prototypes exhibit no significant degradation of their resilience, dissipation, or load transfer capabilities under repeated loading and unloading cycles, as demonstrated in Figures \ref{fig:cyclic}e3, \ref{fig:cyclic}e4, \ref{fig:cyclic}f3 and \ref{fig:cyclic}f4, where $E_{10}/E_1$ = 0.81 and $D_{10}/D_1$ = 0.80 in the maximum modulus direction and $E_{10}/E_1$ = 0.78 and $D_{10}/D_1$ = 0.77 in the minimum modulus direction. Overall, adding connectivity leads to superb reusability, resilience and dissipation in these heterogeneous disordered materials.

\clearpage

\section{Conclusion}
\label{sec:Consclusion}
We have demonstrated, via a combination of micromechanical modeling, experiments and numerical simulations, that heterogeneous disordered materials consisting of elastomeric soft and plastomeric hard domains show extreme resilience and energy dissipation capabilities independent of loading directions. Connectivity throughout the random microstructures has been found to play a key role in achieving nearly isotropic elasticity and inelasticity in these heterogeneous materials. Importantly, we have shown that a small finite number of random spatial points, herein only 7 is needed to achieve the nearly isotropic resilience and energy dissipation capabilities in these materials especially with equal volume fractions of constituents; these findings have been clearly supported by the microstructural features including connectivity (or coordination number), rod length distribution, and local orientational symmetry. Our experimental and numerical results on the 3D-printed prototypes have also shown that, by adding connectivity, all mechanical performance including elastic stiffness, plastic strength, flow stresses, energy dissipation and structural stability has been significantly improved in this new class of heterogeneous disordered materials subjected to a number of repeated cycles. Not only do our results present a new class of heterogeneous disordered materials with isotropic, extreme resilience and dissipation, but they represent an important step towards the harnessing of geometric connectivity throughout the subdomains for isotropy, stability and reusability under harsh mechanical loading scenarios.

Finally, not limited to the mechanically isotropic disordered materials demonstrated in this work, our approach will lead to an emergence of new heterogeneous materials that exhibit outstanding combinations of isotropic physical and chemical functionalities. More specifically, it helps us to design and demonstrate new photonic devices that exhibit not only a complete photonic bandgap (\cite{florescu2009designer,man2013isotropic}) but an isotropic loading bearing capability. Furthermore, it enables the creation of new cellular or porous materials with isotropic heat and mass transport properties, for which the engineered disorder throughout the subdomains plays a crucial role (\cite{zhang2016transport,torquato2018multifunctional,torquato2020predicting,yu2021engineered}).

\section*{Acknowledgment} 
This work is supported by National Research Foundation of Korea (Grant No. RS-2023-00279843).

\renewcommand*\appendixpagename{Appendix}
\renewcommand*\appendixtocname{Appendix}
\begin{appendices}
\numberwithin{equation}{section}
\numberwithin{figure}{section}
\numberwithin{table}{section}

\section{Microstructural details in heterogeneous disordered materials}
\label{Appendix:Microstructural}
\subsection{Representative volume elements for N = 4, 6, 7, 8, 9, 10, 15 and 20}
\label{subsec:RVEs}
\begin{figure}[h]
    \centering
    \includegraphics[width=0.95\textwidth]{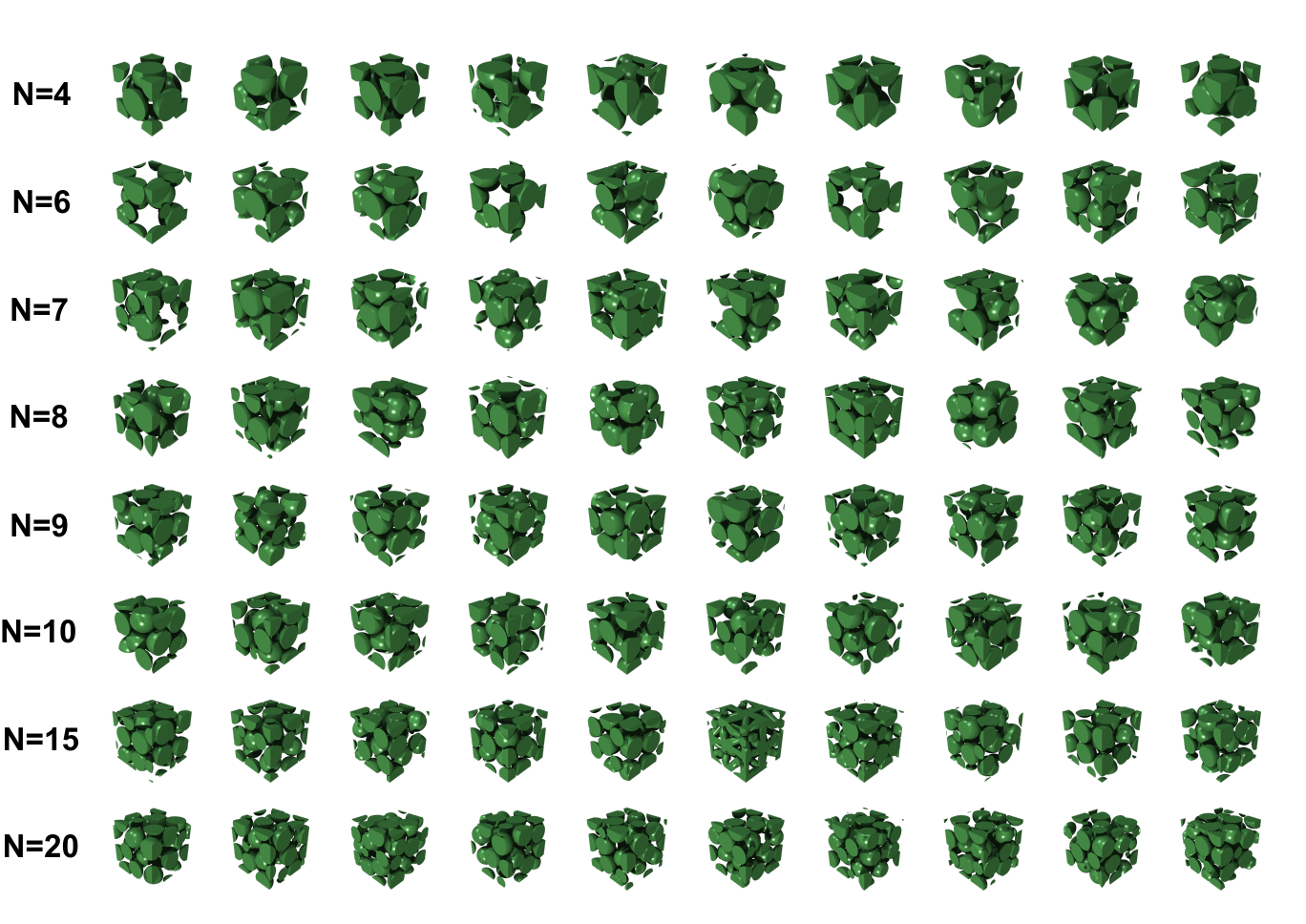}
    \linespread{1.0}
    \caption{Representative volume elements with dispersed-particle morphology (N = 4, 6, 7, 8, 9, 10, 15 and 20). Here, only hard domains are shown (v$_\mathrm{hard}$ = 50\%).}
    \label{fig:rve_disp}
\end{figure}

\begin{figure}
    \centering
    \includegraphics[width=0.95\textwidth]{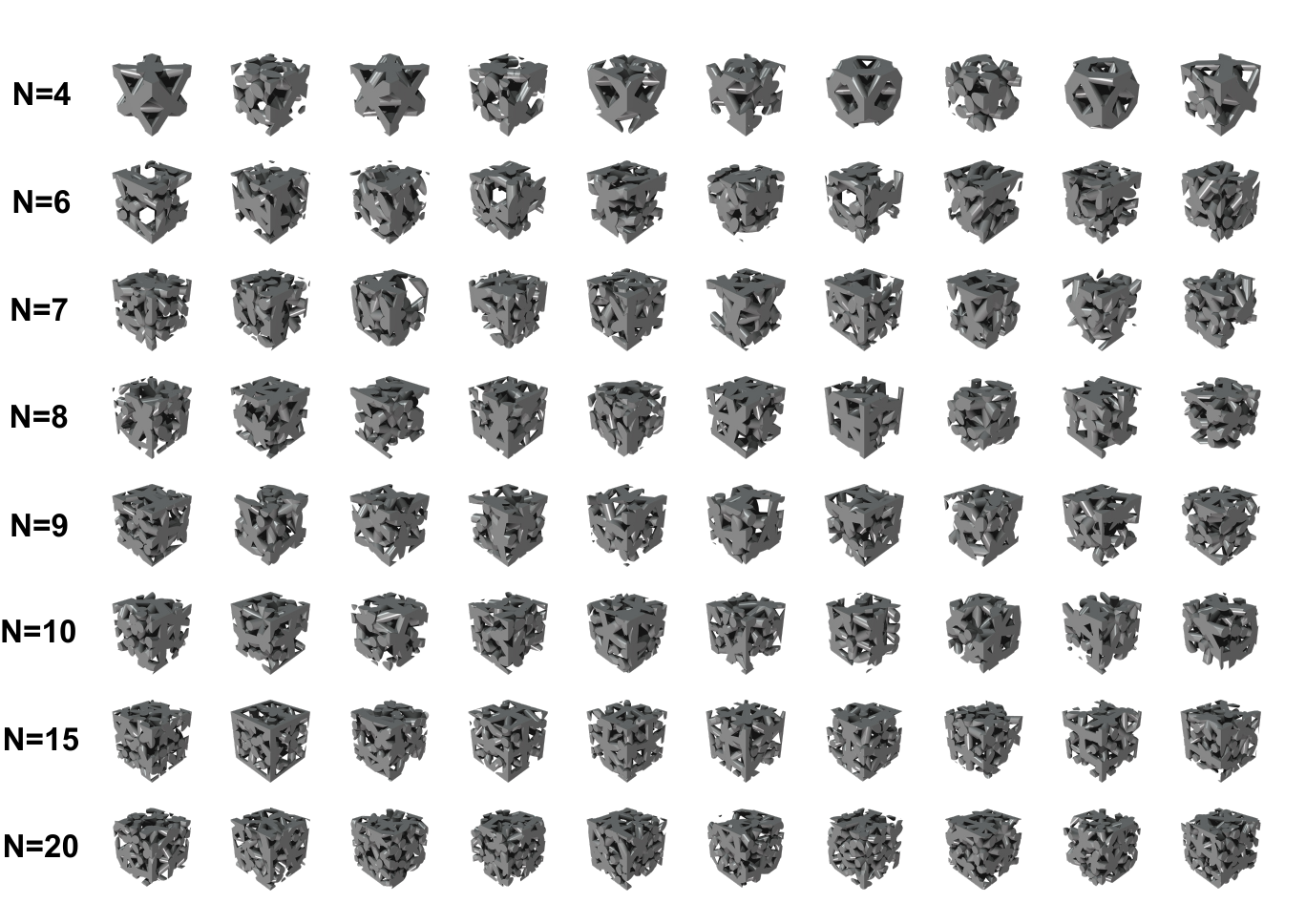}
    \linespread{1.0}
    \caption{Representative volume elements with continuous morphology (N = 4, 6, 7, 8, 9, 10, 15 and 20). Here, only hard domains are shown (v$_\mathrm{hard}$ = 50\%).}
    \label{fig:rve_cont}
\end{figure}

Figures~\ref{fig:rve_disp} and~\ref{fig:rve_cont} show representative volume elements (RVEs) with dispersed-particle and continuous hard domains, which were used throughout this study. In the main text of the manuscript, Figures~\ref{fig:elastic_anisotropy}a and~\ref{fig:elastic_anisotropy}c show the three-dimensional anisotropy maps corresponding to these RVEs. 

\clearpage

\subsection{Maximum achievable packing fractions for N = 2, 3, 4 and 5}
\label{subsec:MaxPackingFraction}
\begin{figure}
    \centering
    \includegraphics[width=0.8\textwidth]{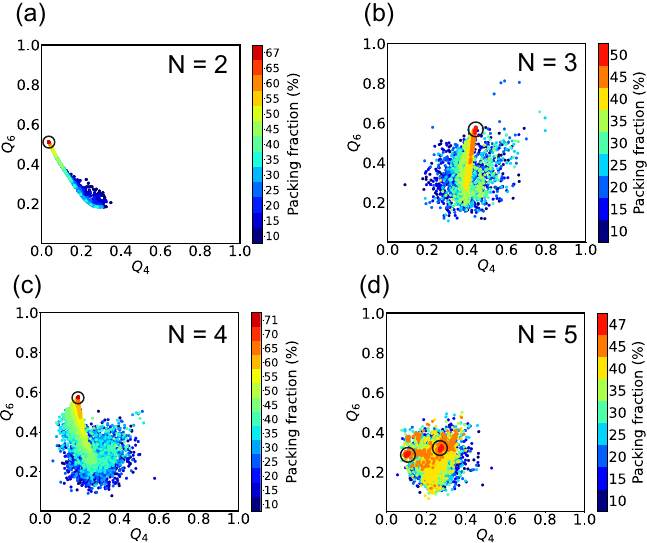}
    \linespread{1.0}
    \caption{Local bond-orientational order parameters ($Q_4$, $Q_6$) in one hundred statistical realizations with (a) N = 2, (b) N = 3, (c) N = 4 and (d) N = 5.}
    \label{fig:q4q6_n2345}
\end{figure}

Figure~\ref{fig:q4q6_n2345} presents the bond-orientational order parameters in one hundred statistical realizations for each of N = 2, 3, 4 and 5 with varying packing fractions. The bond-orientational order quantifies the degree of rotational symmetry in many body systems. Here, we have computed $Q_4$ and $Q_6$ for each of the RVEs. The bond-orientational order parameters are shown to widely disperse with relatively lower packing fractions; as the packing fraction increases, ($Q_4$, $Q_6$) are shown to tend to converge. At the maximum achievable packing fractions (68.02\% for N = 2, 55.56\% for N = 3 and 74.04\% for N = 4), the bond-orientational order parameters are found to converge on ($Q_4$, $Q_6$) = (0.0360, 0.511), (0.446, 0.577) and (0.191, 0.574) for N = 2, 3 and 4, respectively. Interestingly, these values obtained from the random sphere packing are identical to those in body-centered-cubic (bcc), octahedral and face-centered-cubic (fcc) lattices; in other words, the point configurations obtained at the maximum packing fractions for N = 2, 3 and 4 are indeed found to be bcc, octahedral and fcc crystals (Figures~\ref{fig:q4q6_n2345}a,~\ref{fig:q4q6_n2345}b and~\ref{fig:q4q6_n2345}c). Note that a point configuration for N = 1 is simple-cubic under periodic boundary conditions. Similarly, for N = 5, the bond-orientational order parameters are shown to widely disperse at relatively lower packing fractions (v$_\mathrm{pack} < $  45\%), as shown in Figure~\ref{fig:q4q6_n2345}d. At v$_\mathrm{pack}$ $\sim$ 47\%, the bond-orientational order parameters are found to converge on either (0.102, 0.298) or (0.274, 0.318). It shows that the point configurations with N = 5 are locally crystallized with high orientational symmetries at this particular, maximum achievable packing fraction of $\sim$ 47\%. Thus, the simple sphere packing algorithm failed to construct point configurations for N = 5 at v$_\mathrm{pack}$ = 50.75\% (and v$_\mathrm{hard}$ = 50\%) we have explored for these heterogeneous disordered materials in this study.

\clearpage

\subsection{Distributions of lengths of rods in continuous RVEs with N = 6, 7, 8, 9, 10, 15 and 20}
\label{subsec:rod_length}
\begin{figure}[h]
    \centering
    \includegraphics[width=1.0\textwidth]{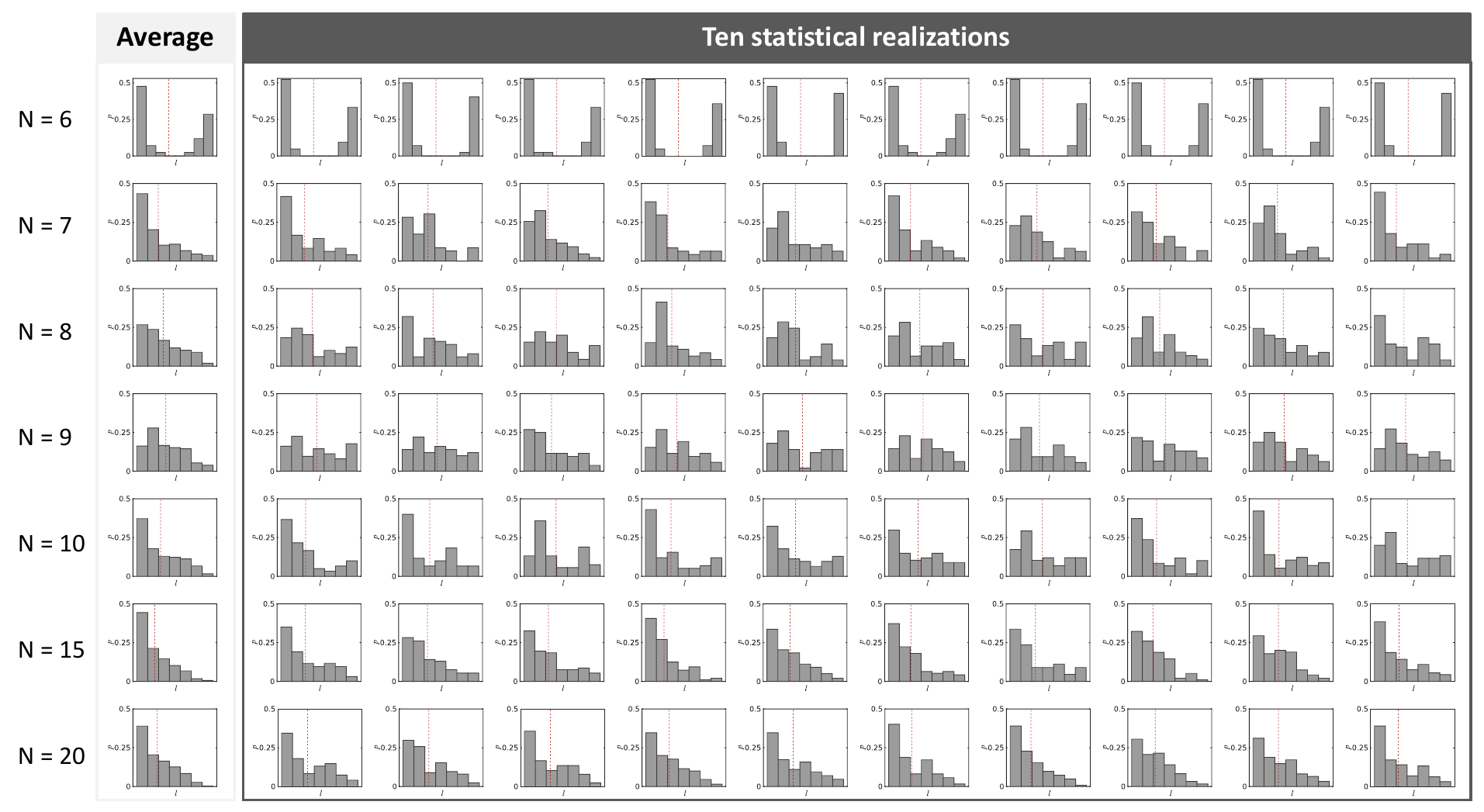}
    \linespread{1.0}
    \caption{Distributions of rod lengths in the RVEs with N = 6, 7, 8, 9, 10, 15 and 20. The red dashed line indicates the average rod length in each distribution.}
    \label{fig:rod_distributions}
\end{figure}

Figure~\ref{fig:rod_distributions} presents the histograms of rod lengths in the continuous RVEs with N = 6, 7, 8, 9, 10, 15 and 20. In the RVEs with N = 6, the distributions of the rod lengths obtained from ten statistical realizations are shown to be bimodal. However, in the RVEs with N = 7, 8, 9, 10, 15 and 20, positively skewed distributions are observed across all of the statistical realizations.

\clearpage

\section{Experimental and numerical procedures}
\label{Appendix:Experimental}
\subsection{Prototype fabrication and mechanical testing}
\label{subsec:fabrication}
A three-dimensional CAD (computer-aided design) program (SolidWorks, Dassault system) was used to prepare the geometries of heterogeneous disordered materials with dispersed-particle and continuous hard domains. Prototypes were then fabricated using a high-precision, multi-material 3D printer (Connex3 Objet260, Stratasys Inc.). Photo-curable pre-polymer droplets were ejected from micro-heads, deposited at specific positions on each printing layer and immediately cured by UV light. In this study, TangoPlus\textsuperscript{TM} (a rubbery polymer) was used for the soft domains and a mixture of TangoPlus\textsuperscript{TM} and VeroWhitePlus\textsuperscript{TM} (a thermoplastic polymer) was used for the hard domains. The stress-strain behavior of each of the hard and soft components is experimentally and numerically characterized in Appendix \ref{subsec:constitutive}. The prototypes have dimensions equivalent to a 5 × 5 × 5 array of unit-cells, with each unit-cell measuring 7 mm per side, resulting in overall dimensions of 35 mm × 35 mm × 35 mm. The radius of cylindrical rods in the RVEs with continuous morphology was determined to meet the desired volume fraction of hard components (here, v$_\mathrm{hard}$ = 50\%). The minimum feature size of the microstructures in each prototype was approximately 1 $\mathrm{mm}$, at least one order of magnitude greater than the printer resolution (15--30 $\mathrm{\mu m}$).

The 3D-printed prototypes were stored away from direct sunlight for 24 hours prior to mechanical testing. We then conducted uniaxial compression tests at a constant engineering strain rate of 0.05 s$^{-1}$ on the prototypes using a testing machine (INSTRON 4482, 100 kN load cell) at room temperature (295 K). Load-displacement data were collected, and a high-resolution digital camera was used to obtain the sequential images of prototypes during deformation. 

\subsection{Mechanical behavior of constituent hard and soft materials}
\label{subsec:constitutive}
\begin{figure}[t]
    \centering
    \includegraphics[width=0.6\textwidth]{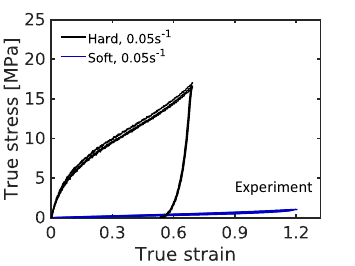}
    \linespread{1.0}
    \caption{Stress-strain behavior of the hard component (black solid line) and the soft component (blue solid line) under uniaxial compression at a strain rate of 0.05 s$^\mathrm{-1}$ in experiments.}
    \label{fig:materials_exp}
\end{figure}

Figure~\ref{fig:materials_exp} presents the stress strain behaviors in the hard (a mixture of VeroWhitePlus\textsuperscript{TM} and TangoPlus\textsuperscript{TM}) and soft (TangoPlus\textsuperscript{TM}) components at an engineering strain rate of 0.05~s$^\mathrm{-1}$. During loading, the hard component exhibits a relatively stiff initial response, followed by a yield-like stress rollover and strain hardening; upon unloading, the hard component displays highly nonlinear behavior with significant energy dissipation and residual strain. In contrast, the soft component shows a much more compliant stress-strain response with negligible energy dissipation and recovers its original shape upon unloading.

\begin{figure}[h]
    \centering
    \includegraphics[width=0.6\textwidth]{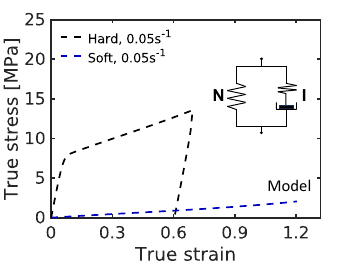}
    \linespread{1.0}
    \caption{Stress-strain behavior of the hard component (black dashed line) and the soft component (blue dashed line) under uniaxial compression at a strain rate of 0.05 s$^\mathrm{-1}$ in numerical simulations.}
    \label{fig:materials_model}
\end{figure}

Constitutive models of the hard and soft components are then presented. The constitutive model of the hard component comprises a time-dependent elastic-inelastic mechanism (denoted I) and a time-independent equilibrium hyperelastic mechanism (denoted N); see the inset of Figure~\ref{fig:materials_model}, which schematically illustrates the microrheological mechanisms in the hard component. We then define the following basic kinematic fields:
\vspace*{0.3in}

\begin{tabular}{l l}
    $\mathbf{F}=\frac{\partial \boldsymbol{\upvarphi}}{\partial \mathbf{X}}=\mathbf{F}_\mathrm{I}=\mathbf{F}_\mathrm{N}$ & deformation gradient that maps material points in reference ($\mathbf{X}$) \\ &to points in deformed spatial configuration ($\mathbf{x} = \boldsymbol{\upvarphi}(\mathbf{X},t)$; $\boldsymbol{\upvarphi}$: motion); \\
    $\mathbf{F}_\mathrm{I} = \mathbf{F}^{e}_\mathrm{I} \mathbf{F}^{p}_\mathrm{I}$ & multiplicative decomposition of $\mathbf{F}_\mathrm{I}$; \\
    $\mathbf{F}^{e}_\mathrm{I} = \mathbf{R}^{e}_\mathrm{I} \mathbf{U}^{e}_\mathrm{I}$ & polar decomposition of $\mathbf{F}^{e}_\mathrm{I}$ into rotation and stretch tensors; \\
    $J = \mathrm{det}(\mathbf{F}) > 0$ & volume change;  \\
    $\bar{\mathbf{F}}_\mathrm{N} = J^{-1/3}\mathbf{F}_\mathrm{N}$ & isochoric part of $\mathbf{F}_\mathrm{N}$;\\
    $\bar{\mathbf{B}}_\mathrm{N} = \bar{\mathbf{F}}_\mathrm{N}\bar{\mathbf{F}}_\mathrm{N}^{\top}$ & isochoric left Cauchy-Green tensor. \\
\end{tabular}
\vspace*{0.3in}

\noindent The deformation rate is described by the velocity gradient $\mathbf{L} = \mathrm{grad}\textbf{v}$, which is decomposed into the elastic and plastic parts,
\begin{equation}
\begin{aligned}
\mathbf{L} & =\dot{\mathbf{F}}\mathbf{F}^{-1} \\
& =\dot{\mathbf{F}}^{e}_{\mathrm{I}}\mathbf{F}^{e-1}_{\mathrm{I}}+\mathbf{F}^{e}_{\mathrm{I}}\dot{\mathbf{F}}^{p}_{\mathrm{I}}\mathbf{F}^{p-1}_{\mathrm{I}}\mathbf{F}^{e-1}_{\mathrm{I}} \\
& = \mathbf{L}^{e}_{\mathrm{I}}+\mathbf{F}^{e}_{\mathrm{I}}\mathbf{L}^{p}_{\mathrm{I}}\mathbf{F}^{e-1}_{\mathrm{I}}.
\end{aligned}
\end{equation}

\noindent Here, the plastic part of the velocity gradient is $\mathbf{L}^{p}_\mathrm{I} = \mathbf{D}^{p}_\mathrm{I} + \mathbf{W}^{p}_\mathrm{I}$, where $\mathbf{D}^{p}_\mathrm{I}$ is the rate of plastic stretching (symmetric part of $\mathbf{L}^{p}_\mathrm{I}$), and $\mathbf{W}^{p}_\mathrm{I}$ is the plastic spin (skew part of $\mathbf{L}^{p}_\mathrm{I}$). In addition, we make two important kinematical assumptions for plastic flow: incompressible flow, i.e., $\operatorname{det}\mathbf{F}^p_\mathrm{I}=1$ and irrotational flow, i.e., $\mathbf{W}^{p}_\mathrm{I}=0$.  Thus, the rate of change in the plastic deformation gradient is given by,
\begin{equation}
\begin{aligned}
        \dot{\mathbf{F}}^p_\mathrm{I} = \mathbf{D}^p_\mathrm{I} \mathbf{F}^{p}_\mathrm{I}.
\end{aligned}
\end{equation}

\noindent The Cauchy stress ($\mathbf{T}_{\mathrm{I}}$) in the time-dependent elastic-inelastic mechanism I is expressed by,
\begin{equation}
\begin{aligned}
\mathbf{T}_{\mathrm{I}} = \frac{1}{J} \mathbf{R}^{e}_{\mathrm{I}} \mathbf{M}^{e}_{\mathrm{I}} \mathbf{R}^{e \top}_{\mathrm{I}}
\end{aligned}
\end{equation}
with the Mandel stress $\mathbf{M}^{e}_{\mathrm{I}} = 2\mu_{\mathrm{I}} (\ln \mathbf{U}^{e}_{\mathrm{I}})_{0} \, + \, K(\ln J)\mathbf{I}$, the deviatoric part of the logarithmic elastic strain $(\ln \mathbf{U}^e_\mathrm{I})_0 = \ln \mathbf{U}^e_\mathrm{I} - \frac{1}{3}\mathrm{tr}(\ln \mathbf{U}^e_\mathrm{I})\mathbf{I}$, the shear modulus $\mu_\mathrm{I}$ and the bulk modulus $K$. We note that the total bulk response is lumped into the mechanism I. The rate of plastic stretching $\mathbf{D}^p_\mathrm{I}$ is assumed to be coaxial to the deviatoric part of the Mandel stress, i.e., $(\mathbf{M}^e_\mathrm{I})_0 = \mathbf{M}^e_\mathrm{I} - \frac{1}{3}\mathrm{tr}(\mathbf{M}^e_\mathrm{I})\mathbf{I}$,
\begin{equation}
\mathbf{D}^{p}_{\mathrm{I}} =\frac{\dot{\gamma}^{p}}{\sqrt{2}}\mathbf{N}^{p}_{\mathrm{I}} \quad \text{where} \quad \mathbf{N}^{p}_{\mathrm{I}} = \frac{(\mathbf{M}^{e}_{\mathrm{I}})_{0}}{\|(\mathbf{M}^{e}_{\mathrm{I}})_{0}\|} \quad \text{and} \quad\|(\mathbf{M}^{e}_{\mathrm{I}})_{0}\| = \sqrt{(\mathbf{M}^{e}_{\mathrm{I}})_{0} : (\mathbf{M}^{e}_{\mathrm{I}})_{0}}.
\end{equation}
Then, we employed the thermally-activated viscoplasticity model prescribed by,
\begin{equation}
    \dot{\gamma}^{p} = \dot{\gamma}_{\mathrm{0}} \operatorname{exp}\left[ -\frac{\Delta G}{k\theta} \left\{1-\frac{\bar{\tau}}{s_{0}}  \right\}   \right] \quad \text{where} \quad \bar{\tau} = \frac{1}{\sqrt{2}} \|(\mathbf{M}^{e}_{\mathrm{I}})_{0}\|,
\end{equation}
with the reference plastic strain rate $\dot{\gamma}_{\mathrm{0}}$, the activation energy $\Delta G$, Boltzmann's constant $k$, the absolute temperature $\theta = 295\ \mathrm{K}$ (room temperature) and the shear strength $s_{0}$. Additionally, $\bar{\tau}$ is the magnitude of the deviatoric Mandel stress.

The Cauchy stress ($\mathbf{T}_{\mathrm{N}}$) in the time-independent mechanism N is given by,
\begin{equation}
\mathbf{T}_{\mathrm{N}} = \frac{\mu_{\mathrm{N}}}{3J} \frac{\lambda_{\mathrm{N}}}{\bar{\lambda}} \mathscr{L}^{-1} \left(\frac{\bar{\lambda}}{\lambda_{\mathrm{N}}}\right) (\bar{\mathbf{B}}_{\mathrm{N}})_0 \quad \text{where} \quad \bar{\lambda} = \sqrt{\frac{\text{tr}\bar{\mathbf{B}}_{\mathrm{N}}}{3}} \quad \text{and} \quad (\bar{\mathbf{B}}_{\mathrm{N}})_0 = \bar{\mathbf{B}}_{\mathrm{N}} - \frac{1}{3}\mathrm{tr}(\bar{\mathbf{B}}_{\mathrm{N}})\mathbf{I},
\end{equation}
with the shear modulus $\mu_{\mathrm{N}}$ and the limiting chain extensibility $\lambda_{\mathrm{N}}$. Also, $\mathscr{L}^{-1}$ is the inverse Langevin function, $\mathscr{L}(x)=\coth(x)-\dfrac{1}{x}$.

The total stress in the hard component is obtained by,
\begin{equation}
\mathbf{T}_{\mathrm{hard}} = \mathbf{T}_{\mathrm{I}} + \mathbf{T}_{\mathrm{N}}.
\end{equation}

For the soft component, we have used a nearly incompressible Arruda-Boyce model (\cite{arruda1993three}). The Cauchy stress is expressed by,
\begin{equation}
\begin{aligned}
    \mathbf{T}_{\mathrm{soft}}=\frac{\mu_\mathrm{soft}}{3 J} \frac{\lambda_\mathrm{soft}}{\bar{\lambda}} \mathscr{L}^{-1}\left(\frac{\bar{\lambda}}{\lambda_\mathrm{soft}}\right)\left(\bar{\mathbf{B}}\right)_0 + {K_\mathrm{soft}} (J-1) \mathbf{I} \quad &\text{where} \quad \bar{\lambda} = \sqrt{\frac{\mathrm{tr}(\bar{\mathbf{B}})}{3}} \\ &\text{and} \quad \bar{\mathbf{B}} = J^{-2/3} \mathbf{F} \mathbf{F}^\top,
\end{aligned}
\end{equation}
where $\mu_\mathrm{soft}$ is the shear modulus, $K_\mathrm{soft}$ is the bulk modulus, $\lambda_\mathrm{soft}$ is the limiting chain extensibility in soft components, $(\bar{\mathbf{B}})_0 = \bar{\mathbf{B}} - \frac{1}{3}\mathrm{tr}(\bar{\mathbf{B}})\mathbf{I}$ is the deviatoric part of the isochoric left Cauchy-Green tensor $\bar{\mathbf{B}}$, $\mathbf{F}$ is the deformation gradient and $J = \mathrm{det}(\mathbf{F})$ is the volume change in the soft component.

Figure~\ref{fig:materials_model} presents the stress-strain curves numerically simulated using the constitutive models for both hard and soft components. These models are found to reasonably capture the overall plastomeric and elastomeric features in the constituent materials during a loading and unloading cycle. The material parameters used in the models are provided in Table~\ref{Tab:material_parameter}. The finite deformation constitutive model for the hard component was then numerically implemented for use in the finite element solver (Abaqus/Standard) while an approximate Arruda-Boyce hyperelastic representation available in Abaqus/Standard was used for the soft component. For detailed information on the time integration procedures for the time-dependent mechanism (I), see the work of \cite{weber1990finite}.

\begin{table}[htb]
\centering
\small{
\begin{tabular}{lcc}
\hline
\textbf{Hard component: Time-dependent elastic-inelastic mechanism I} &  &  \\
\hline
& $\mu_{\mathrm{I}}$ [MPa] & 42.77 \\
$\mathbf{T}_{\mathrm{I}} = \frac{1}{J} \mathbf{R}^{e}_{\mathrm{I}} \mathbf{M}^{e}_{\mathrm{I}} \mathbf{R}^{e \top}_{\mathrm{I}} \quad \text{where} \quad \mathbf{M}^{e}_{\mathrm{I}} = 2\mu_{\mathrm{I}} (\ln \mathbf{U}^{e}_{\mathrm{I}})_{0} \, + \, K(\ln J)\mathbf{I}$ & $K$ [GPa] & 0.55 \\
& $\Delta G$ [$10^{-20}$J] & 1.49 \\
$\dot{\gamma}^{p} = \dot{\gamma}_{\mathrm{0}} \operatorname{exp}\left[ -\frac{\Delta G}{k\theta} \left\{1-\frac{\bar{\tau}}{s_{0}}  \right\}   \right] \quad \text{where} \quad \bar{\tau} = \frac{1}{\sqrt{2}} \|(\mathbf{M}^{e}_{\mathrm{I}})_{0}\|$ & $\dot{\gamma}_{\mathrm{0}}$ [s$^{-1}$] & 0.025 \\
& $s_{0}$ [MPa] & 3.099 \\
\hline
\textbf{Hard component: Time-independent hyperelastic network N} &  &   \\
\hline
$\mathbf{T}_{\mathrm{N}} = \frac{\mu_{\mathrm{N}}}{3J} \frac{\lambda_{\mathrm{N}}}{\bar{\lambda}} \mathscr{L}^{-1} \left(\frac{\bar{\lambda}}{\lambda_{\mathrm{N}}}\right) (\bar{\mathbf{B}}_{\mathrm{N}})_0$ & $\mu_{\mathrm{N}}$ [MPa] & 2.52 \\
& $\lambda_{\mathrm{N}}=\sqrt{N_\mathrm{N}}$ & $\sqrt{6}$ \\
\hline
\textbf{Soft component} &  &   \\
\hline
& $\mu_{\mathrm{soft}}$ [MPa] & 0.53  \\
$ \mathbf{T}_{\mathrm{soft}}=\frac{\mu_\mathrm{soft}}{3 J} \frac{\lambda_\mathrm{soft}}{\bar{\lambda}} \mathscr{L}^{-1}\left(\frac{\bar{\lambda}}{\lambda_\mathrm{soft}}\right)\left(\bar{\mathbf{B}}\right)_0 + {K_\mathrm{soft}} (J-1) \mathbf{I}$ & $K_\mathrm{soft}$ [MPa] & 26.32  \\
& $\lambda_\mathrm{soft}=\sqrt{N_\mathrm{soft}}$ & $\sqrt{10}$ \\
\hline
\end{tabular}
}
\caption{Material parameters used in the constitutive models for the hard and soft components.}
\label{Tab:material_parameter}
\end{table}
\clearpage

\subsection{Micromechanical modeling}
\label{subsec:Micromechanical}
The boundary value problems for the RVEs with both morphologies were solved using nonlinear finite element analysis with Abaqus/Standard. To this end, the finite deformation constitutive models for the hard and soft components presented in Appendix \ref{subsec:constitutive} were numerically implemented for use in the nonlinear finite element solver. The macroscopic average responses of the RVEs were computed under periodic boundary conditions (PBC) using the fictitious node virtual work method (\cite{danielsson2002three,danielsson2007micromechanics}). Additionally, we used quadratic elements in the finite element analysis.

A generalized anisotropic linear elasticity is described by a stress ($\mathbf{T}$) - strain ($\mathbf{E}$) relationship of the form of $\mathbf{T} = \mathbb{C} \mathbf{E}$, where $\mathbb{C}$ is the fourth-order stiffness tensor. In fully anisotropic materials (i.e., with no symmetries), the stiffness tensor has 21 independent elastic moduli components and can be represented in a symmetric 6 × 6 matrix using a Voigt form. To determine these independent elastic components, we solved boundary value problems of the RVEs subjected to six independent macroscopic deformation conditions (three uniaxial compressions and three simple shears). Once the fourth-order stiffness tensor is obtained for a given RVE, the directional elastic moduli with respect to arbitrary crystallographic orientations can be computed (e.g., see Figures \ref{fig:elastic_anisotropy}a and \ref{fig:elastic_anisotropy}c).

Then, the universal anisotropy index ($\mathrm{A}^\mathrm{U}$) used throughout the main body of this work is calculated as follows (e.g., see Figures \ref{fig:elastic_anisotropy}b and \ref{fig:elastic_anisotropy}d) (\cite{ranganathan2008universal}):
\begin{equation}
    \mathrm{A}^\mathrm{U} = 5 \frac{G^V}{G^R} + \frac{K^V}{K^R} - 6 \geq 0,
\end{equation}
where $G$ and $K$ are the shear modulus and bulk modulus, respectively, and the superscripts $V$ and $R$ represent the Voigt and Reuss bounds. Note that the universal anisotropy $\mathrm{A}^\mathrm{U}$ is greater than or equal to zero, and for an isotropic elastic material, $\mathrm{A}^\mathrm{U} = 0$. The equivalent Zener index ($\mathrm{A}^\mathrm{eq}$) is also calculated by (\cite{nye1985physical, ranganathan2008universal}),
\begin{equation}
    \mathrm{A}^\mathrm{eq} = (1 + \frac{5}{12} \mathrm{A}^\mathrm{U}) + \sqrt{(1 + \frac{5}{12} \mathrm{A}^\mathrm{U})^2 - 1}.
\end{equation}

\section{Further analysis on inelastic isotropy in continuous RVE}
\label{sec:inelastic_isotropy}
\begin{figure}[h]
    \centering
    \includegraphics[width=0.8\textwidth]{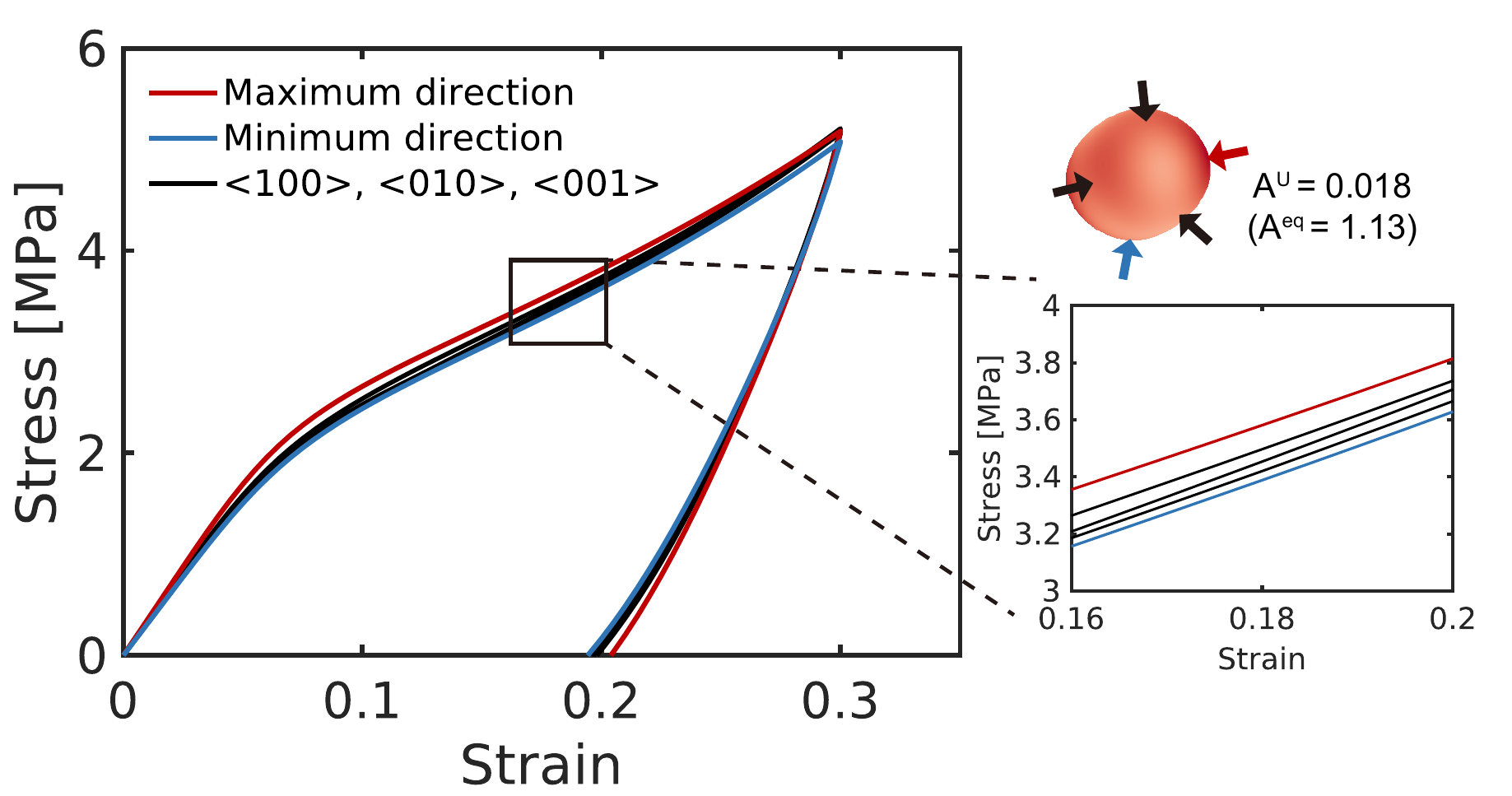}
    \linespread{1.0}
    \caption{Micromechanical modeling results for a continuous RVE with N = 7 (v$_\mathrm{hard}$ = 50\%); red line: macroscopic stress-strain curve in maximum modulus direction, blue line: macroscopic stress-strain curve in minimum modulus direction, black lines: macroscopic stress-strain curves in $<$100$>$, $<$010$>$ and $<$001$>$ directions, apparently lied between those in the maximum and minimum modulus directions.}
    \label{fig:inelastic}
\end{figure}

Figure~\ref{fig:inelastic} presents micromechanical modeling results for representative disordered RVE with continuous morphology (N = 7, $\mathrm{A}^{\mathrm{U}}$ = 0.018), loaded in its maximum and minimum modulus directions, as well as in the $<$100$>$, $<$010$>$ and $<$001$>$ directions. The initial elastic modulus, inelastic yield, flow stresses, energy dissipation and residual strain upon unloading are found to be nearly identical across all loading directions.

\clearpage

\section{List of supplementary movies}
\label{sec:Movies}
All movies are available online in the following repository. \\
\href{https://solidslabkaist.github.io/archives/movies/heterogeneous_disordered/}{https://solidslabkaist.github.io/archives/movies/heterogeneous\_disordered/}

\begingroup
\linespread{1.5}\selectfont

\paragraph{Movie S1}
The experimental images are presented together with the macroscopic stress-strain curves of a 3D-printed prototype with dispersed-particle morphology, loaded in its maximum modulus direction. A highly resilient and energy-dissipative behavior is observed during the multiple cycles of loading, unloading and recovery; the load transfer and energy dissipation capabilities decrease significantly in the subsequent cycles. Furthermore, buckling throughout the hard particles, evidenced by the second drop in the instantaneous tangle modulus becomes more pronounced in the subsequent cycles. See Figure~\ref{fig:cyclic}c for more details.

\paragraph{Movie S2}
The experimental images are presented together with the macroscopic stress-strain curves of a 3D-printed prototype with dispersed-particle morphology, loaded in its minimum modulus direction. Instability in this minimum modulus loading direction is shown to be mitigated in the subsequent cycles of N$_\mathrm{C}$ = 2 to N$_\mathrm{C}$ = 10. Though the prototype is shown to be stable, the load transfer and energy dissipation capabilities significantly degrade during the multiple cycles of N$_\mathrm{C}$ = 1 to N$_\mathrm{C}$ = 10, similarly to the prototype with dispersed-particle morphology loaded in its maximum modulus direction. See Figure~\ref{fig:cyclic}d for more details.

\paragraph{Movie S3}
The experimental images are presented together with the macroscopic stress-strain curves of a 3D-printed prototype with continuous hard domains, loaded in its maximum modulus direction. Highly stable deformation is observed without any buckling throughout the hard ligament network during multiple cycles. Furthermore, the prototype exhibits no significant degradation in the resilience, dissipation and load transfer capabilities under repeated loading, unloading, recovery and reloading cycles. See~Figure~\ref{fig:cyclic}e for more details.

\paragraph{Movie S4}
The experimental images are presented together with the macroscopic stress-strain curves of a 3D-printed prototype with continuous hard domains, loaded in its minimum modulus direction. The stress-strain curves in this minimum modulus direction are nearly identical to those in the maximum elastic modulus direction, thus revealing nearly isotropic elastic and inelastic features in these materials with continuous morphology. Furthermore, the highly stable response without any significant degradation in resilience, dissipation and load transfer is observed during N$_\mathrm{C}$ = 1 to N$_\mathrm{C}$ = 10. See Figures~\ref{fig:cyclic}e and~\ref{fig:cyclic}f for more details.

\endgroup
\end{appendices}
\clearpage
\printbibliography
\end{document}